\input phyzzx
\vsize 9.25in
\hsize 6.3in
\def\scri{{\cal I}}
\def\hg{\hat g}
\def\hR{\hat R}
\def\hn{\hat \nabla}
\def\vf{\vec f}
\def\cR{\cal R}
\rightline{UATP-96/01}
\rightline{March 1996}
\vskip 0.2in 
\centerline{\seventeenbf Quantum Naked Singularities}
\centerline{\seventeenbf in 2d Dilaton Gravity}
\vskip 0.5in
\centerline{\caps Cenalo Vaz\footnote{\dagger}{Internet:
cvaz@mozart.si.ualg.pt}}
\centerline{\it Unidade de Ci\^encias Exactas e Humanas}
\centerline{\it Universidade do Algarve}
\centerline{\it Campus de Gambelas, P-8000 Faro, Portugal}
\vskip 0.2in
\centerline{and}
\vskip 0.2in
\centerline{\caps Louis Witten\footnote{\dagger\dagger}{Internet:
witten@physunc.phy.uc.edu}}
\centerline{\it Department of Physics}
\centerline{\it University of Cincinnati}
\centerline{\it Cincinnati, OH 45221-0011, U.S.A.}
\vskip 0.75in
\centerline{\bf \caps Abstract}
\vskip 0.2in

\noindent Roughly speaking, naked singularities are singularities
that may be seen by timelike observers. The Cosmic Censorship
conjecture forbids their existence by stating that a reasonable
system of energy will not, under reasonable conditions, collapse
into a naked singularity. There are however many (classical)
counter-examples to this conjecture in the literature. We propose
a defense of the conjecture through the quantum theory. We will
show that the Hawking effect and the accompanying back reaction,
when consistently applied to naked singularities in two dimensional
models of dilaton gravity with matter and a cosmological constant,
prevent their formation by causing them to explode or to emit
radiation catastrophically. This contrasts with black holes which
radiate slowly. If this phenomenon is reproduced in the four
dimensional world, the radiation accompanying the formation of
naked singularities should have observable consequences.
\vfill
\eject

\noindent {\bf I. Introduction}

Two dimensional models of gravity seem to mimic the behavior of
their four dimensional counterparts. They have therefore long been
considered excellent theoretical laboratories for the examination
of the various conceptual problems posed by Hawking's original work
on black hole evaporation.${}^1$ One important advantage of two
dimensions is in the fact that the quantum stress energy tensor of
a conformal field propagating in a curved two dimensional
background can, in most cases, be calculated exactly from a
knowledge of the trace anomaly. Thus, for example, Christensen and
Fulling${}^2$ were able to obtain the exact stress tensor of
evaporation of a two dimensional Schwarzschild black hole,
neglecting the back reaction of the spacetime, and their stress
tensor exhibited most of the physical properties expected of the
full four dimensional theory.

More recent interest in two dimensional models has been spurred by
Witten's discovery of an exact conformal field theory describing
dilaton gravity in two dimensions.${}^3$ Witten's model also arises
from the compactification on an eight torus of ten dimensional
dilaton gravity originating in heterotic string theory to lowest
order in world-sheet perturbation theory,${}^4$ and as the
effective action describing the radial modes of extreme dilatonic
black holes in four dimensions.${}^5$ It possesses a rich structure
with timelike and spacelike singularities and relatively simple
dynamics, while reproducing most of the key features of four
dimensional Einstein gravity. The model has been exploited,
therefore, to revisit some of the fundamental problems raised by
Hawking's early work. As a consequence, quantum black holes in two
dimensions have received, and are receiving, a good deal of
attention.${}^{6,7,8,9}$ 

An important model of dynamical formation and evaporation of a
black hole in two dimensional dilaton gravity was proposed by
Callan, Giddings, Harvey and Strominger (CGHS).${}^6$ The authors
coupled conformal matter degrees of freedom to the original Witten
model and then examined an incoming shock wave of the matter
fields, finding that it can be expected to radiate away all its
energy before the black hole is actually formed. There were
difficulties with this conclusion, however, because the back
reaction of the spacetime was neglected as a first approximation
and later could not be adequately taken into account. The dilaton
coupling was also seen to be too large at the turn around point,
further belying the original one loop calculation. Since this
proposal, various generalizations and improvements have been made
to the CGHS model${}^{7,8,9}$ and the evaporation of a black hole
via particle production of massless conformal matter and including
the back reaction of the geometry of spacetime can be traced
numerically and, in some models, analytically.

Naked singularities,${}^{10}$ on the other hand, have been for the
most part ignored. Roughly speaking naked singularities are
singularities that may be seen by physically allowed observers and
are known to be formed under fairly generic conditions in four
dimensions${}^{11}$. Nevertheless, they have so far been considered
intuitively undesirable and this has produced a conjecture
forbidding their existence, the so-called cosmic censorship
conjecture.${}^{12}$ The conjecture simply states that a physically
``reasonable'' system of energy will not, under ``reasonable''
initial conditions, evolve into a naked singularity. The original
intent of the conjecture was of course to prohibit the formation of
naked singularities in classical general relativity so that, for
example, a non-rotating system would radiate away all multipole
moments sufficiently rapidly that the final state would be a
Schwarzschild black hole. However, there are many counter examples
in the literature where naked singularities do indeed form from
classically reasonable initial conditions.

The existence of naked singularities is still difficult to
understand physically, and several suggestions have been made to
maintain the viability of the cosmic censorship hypothesis. One is
merely to reiterate that it holds classically but needs some fine
tuning to describe the set of physically reasonable set of states
from which a collapse may begin. The fine tuning would be such as
to exclude all known examples of the formation of naked
singularities. A second way to preserve the conjecture is to argue
from the point of view of string theory. String theory is said to
solve the short distance problems of general relativity by
providing a fundamental length scale proportional to the inverse
square root of the string tension. The expectation is therefore
that the high energy behavior of string theory (in weak coupling)
would forbid any observations of processes associated with high
field gradients over short distances. Yet a third possibility,
which we argue for here, is that the Hawking radiation and the
accompanying back reaction will be so huge as to prevent a naked
singularity from forming. 

In ref.[13] we  examined the classical formation and quantum
evaporation of naked singularities within the context of the CGHS
shock-wave model. Our preliminary work seemed to indicate that
naked singularities ``explode'' (as opposed to black holes that are
expected to ``evaporate'') as soon as they are formed. What we mean
by this is that, in the regime in which the dilaton coupling is
small, asymptotically timelike singularities were observed to give
up all their energy in a burst upon forming, which contrasts with
the expectation of relatively slow evaporation from the black hole
(except possibly in its final stage). Subsequent analysis of other
models seems to verify this conclusion, if the naked singularities
actually form classically. It seems possible, as we will show
below, that initial conditions leading to the formation of a naked
singularity are quantum mechanically inconsistent. Therefore naked
singularities will not form. Thus, what we will describe below is
evidence in favor of a cosmic censorship, one with its origins in
the quantum theory.

The purpose of this article is to put together some of the key
results, making a coherent case for the preservation of the cosmic
censorship hypothesis by the quantum theory. We will consider two
dimensional models that are described by an action of the form
$$S~~ =~~ \int d^2x {\sqrt{-g}} \left[ e^{-2\phi} \left( - R~ +~ 4
(\nabla \phi)^2~ +~ \Lambda \right)~ -~ {1 \over 2} \sum_i (\nabla
f_i )^2~ +~ 4\mu^2 e^{-2 \phi} U(f_i) \right], \eqno(1.1)$$
where $U(f_i)$ is a consistent potential ascribed to the scalar
fields. Consistency will mean the preservation of conformal
invariance; this must be checked in the limit of weak dilaton
coupling.${}^7$ We will show that the models classically admit both
timelike and spacelike singularities, their behavior at infinity
depending on the sign of the cosmological constant. They will then
be analyzed in the semi-classical approximation. We first consider,
in section II, the original CGHS model ($U(f_i) = 0$), but with a
negative cosmological constant. A classical naked singularity
appears, instead of a black hole, and we will argue that the naked
singularity is unstable quantum mechanically, giving up all its
energy in a burst so that an asymptotic observer sees only a burst
of thermal radiation. In this model, the naked singularity is not
``formed'' in the sense of being in the future of the infalling
matter. Instead, the matter shock wave and the singularity appear
simultaneously in the spacetime. 

To preserve scale and diffeomorphism invariance in the conformal
gauge, the kinetic terms and the potential terms should get
renormalized so that the full theory becomes a conformal field
theory.${}^7$ In section III we consider what constraints must be
imposed on the potential term, $U(f_i)$, so that the resulting
theory is indeed a consistently quantizable conformal field theory.
We will derive a simple consistency condition and present a
solution. We will see that the sine-Gordon potential is a
particular case of this solution and will go on to analyze the
singularities induced by sine-Gordon solitons${}^{14}$ in section
IV. The latter singularities are interesting even on the classical
level. They are neither purely spacelike nor are they purely
timelike, being ``hybrids'' in the sense that they are timelike in
some regions and spacelike in others: we will show that the nature
of the Hawking radiation depends crucially on the asymptotic
behavior of the singularities. Asymptotically spacelike
singularities will ``evaporate'' while  asymptotically timelike
singularities, if they form, prefer to rapidly radiate their energy
away at early times, or to ``explode''.  In section V we will
examine a particularly interesting singularity consisting of two
timelike pieces that are joined smoothly in the distant future by
a spacelike line. This is an interesting object because the
singularity is formed in the future of the infalling matter, and
thus it represents a possible collapse scenario. However, we will
hold that what seem to be natural initial conditions (no incoming
flux of energy across $\scri^-$ except for the soliton) leads to a
violation of the weak energy conditions if quantum effects are
accounted for. On the other hand requiring  that the weak energy
condition is maintained in the semi-classical theory implies that
the incoming soliton is necessarily accompanied by an incoming flux
of Hawking energy. The effect on the history of the collapse of
this change in initial conditions is examined in section VI which
deals with the back reaction of the spacetime in the context of the
full quantum theory. In this section we examine in greater detail
two of the naked singularities described before, viz., the shock
wave induced naked singularity of section II and the singularity
formed in the future of the incoming soliton of the previous
section. In the first case, that of the shock wave, the full
quantum theory admits no positive mass singularities, spacelike or
timelike, if the Hawking boundary conditions are imposed. Thus, in
this model, quantum effects do prevent the formation of a naked
singularity. In the second case, the two sets of boundary
conditions described above are separately examined.  When we
require that there is no incoming Hawking radiation but permit a
violation of the positive energy conditions, the naked singularity
still exists but has changed its character and location. If we
demand that the semi-classical theory has no violation of the
positive energy conditions we are forced to allow incoming Hawking
radiation and the singularity no longer forms.  We conclude in
section VII with a brief discussion of the implications of these 
results. In what follows our conventions are those of
Weinberg${}^{15}$
\vskip 0.25in

\noindent{\bf II. Shock wave induced naked singularity}

Consider the action
$$S~~ =~~ \int d^2x {\sqrt{-g}} \left[ e^{-2\phi} \left( - R~ +~ 4
(\nabla \phi)^2~ +~ \Lambda \right)~ -~ {1 \over 2} \sum_i (\nabla
f_i )^2 \right] \eqno(2.1)$$
where $\Lambda$ is the cosmological constant, $f_i (x)$ are $N$
conformally coupled matter fields and $\phi$ is the dilaton. It is
the potential free version of (1.1).

The model with a positive cosmological constant, $\Lambda = + 4
\lambda^2$ was initially analyzed by CGHS and shown to be unstable
against gravitational collapse, admitting black hole solutions
produced by incoming $f$ shock waves. It was then shown that
quantum effects appeared to cause the $f-$wave to radiate away its
energy before the formation of the horizon, so that it would seem
that black hole states are excluded from the quantum spectrum. CGHS
simultaneously pointed out several problems (some of which have
been mentioned in the introduction) associated with this na\"\i ve
conclusion and attempts were subsequently made not simply to
eliminate these obstacles but also to construct some exactly
solvable models of black hole evaporation.${}^{8,9,16}$

We will now see that the model with a {\it negative} cosmological
constant, $\Lambda = -4\lambda^2$ yields a naked singularity when
the spacetime is perturbed by an incoming $f$ shock wave. As usual,
the metric equations of motion,
$$\eqalign{0~~ =~~ {\cal T}_{\mu\nu}~~ &=~~ e^{-2\phi}~ \left[ 2
\nabla_\mu \nabla_\nu \phi~ -~ {1 \over 2} e^{2\phi} \sum_i
\nabla_\mu f_i \nabla_\nu f_i \right. \cr &~~~~~~~~ \left. +~~
g_{\mu\nu}~ \left( -2 \nabla^2 \phi~ +~ 2 (\nabla \phi)^2~ +~
2\lambda^2~ +~ {1 \over 4} e^{2\phi} \sum_ i (\nabla f_i)^2 \right)
\right], \cr} \eqno(2.2)$$
where $\nabla_\mu$ is the covariant derivative compatible with
$g_{\mu\nu}$, form a set of constraints on the allowable solutions
of the field equations, viz.,
$$\eqalign{4 \nabla^2 \phi~ -~ 4 (\nabla \phi)^2~ -~ R~ -~
4\lambda^2 ~~ &=~~ 0\cr -\nabla^2 f_i~~ &=~~ 0.\cr}\eqno(2.3)$$
The system is best analyzed in the conformal gauge, $g_{\mu\nu} =
e^{2\rho} \eta_{\mu\nu}$, where $\eta_{\mu\nu}$ is the Lorentz
metric in two dimensions. We will also use light-cone coordinates,
$x^\pm = x^0 \pm x^1$ where the constraint equations become
$$\eqalign{{\cal T}_{++}~~ &=~~ e^{-2\phi} \left[ 2 \partial_+
\partial_+ \phi~ -~ 4 \partial_+ \phi \partial_+ \rho \right]~ -~
{1 \over 2}~ \sum_i \partial_+ f_i \partial_+ f_i~~ =~~ 0 \cr {\cal
T}_{--}~~ &=~~ e^{-2\phi} \left[ 2 \partial_- \partial_-\phi~ -~ 4
\partial_- \phi \partial_- \rho \right]~ -~ {1 \over 2}~ \sum_i
\partial_- f_i \partial_- f_i~~ =~~ 0 \cr {\cal T}_{+-}~~ &=~~ e^{-
2\phi} \left[ -2 \partial_+ \partial_- \phi~ +~ 4 \partial_+ \phi
\partial_- \phi~ -~ \lambda^2 e^{2\rho} \right]~~ =~~ 0
\cr}\eqno(2.4)$$
and the equations of motion are
$$\eqalign{& -~ 4\partial_+ \partial_- \phi~ +~ 4 \partial_+ \phi
\partial_- \phi~ +~ 2 \partial_+ \partial_- \rho~ -~ \lambda^2
e^{2\rho}~~ =~~ 0 \cr & +~ \partial_+ \partial_- f_i~~ =~~ 0. \cr}
\eqno(2.5)$$
Combining the first equation in (2.5) with the last in (2.4) shows
that $\rho(x)$ is the same as $\phi(x)$ up to a harmonic function
$h(x)$. Because the conformal gauge does not fix the conformal
subgroup of diffeomorphisms, a choice of $h(x)$ is essentially a
choice of coordinate system. If we work in the Kruskal gauge, with
$h(x) = 0$, then the solution to the equations of motion,
satisfying all the constraints (with $f_i(x) = 0$) is unique up to
an additive constant,
$$e^{-2\phi}~~ =~~ e^{-2\rho}~~ =~~ \sigma~~ =~~ \lambda^2 x^+ x^-
~~ +~~ {M \over \lambda}.\eqno(2.6)$$
We show below that it represents a naked singularity with Bondi
mass $M$ which we take to be greater than zero. When $M$ is zero
the spacetime is flat and the dilaton is linear in the spatial
coordinate so that this solution is generally referred to as the
linear dilaton vacuum. On the other hand, when $M \neq 0$, the
curvature is given by
$$\eqalign{R~~ &=~~ +~ 2 e^{-3\rho} \nabla^2 e^\rho~~ -~~ 2 e^{-
4\rho} (\nabla e^\rho)^2 \cr &=~~ +~ 4~ \left[ \partial_+
\partial_-\sigma ~~ -~~ {{\partial_+ \sigma \partial_- \sigma}
\over \sigma} \right] \cr &=~~ +~ {{4 \lambda M} \over \sigma}.
\cr} \eqno(2.7)$$
It is singular at $\sigma(x) = 0$ and the singularity is timelike.
Its Penrose diagram is shown in figure I, in which the section of
spacetime with $\sigma > 0$ is labeled I and II. 

This spacetime admits a Killing vector ${\vec \xi}$ given by
$$(\xi^+, \xi^-)~~ =~~ \lambda (x^+, -x^-) \eqno(2.8)$$
which is spacelike in region I of the diagram and timelike in
region II. The ADM mass itself has no meaning, as spatial infinity
is cut off by the singularity, and in such cases one seeks to
define the mass, not at $i^0$ but along $\scri^+$ as follows. The
existence of the Killing vector implies the existence of a
conserved current 
$$j_\mu~~ =~~ {\cal T}_{\mu\nu} \xi^\nu.\eqno(2.9)$$
Let $t_{\mu\nu}$ be a linearizaton of ${\cal T}_{\mu\nu}$ about the
dilaton vacuum so that, to first order, $j_\mu = t_{\mu\nu}
\xi^\nu$ and consider the solution of the dilaton field that is
asymptotic to the vacuum:
$$\eqalign{\phi(x)~~ &=~~ \phi^{(0)}(x)~~ +~~ \delta \phi(x)\cr
\phi^{(0)}~~ &=~~ -~{ 1 \over 2}~ \ln (\lambda^2 x^+ x^-)\cr}
\eqno(2.10)$$
so that the current in (2.9) takes the form
$$\eqalign{j_+~~ &=~~ 2 \lambda \partial_+ \left( e^{-2\phi^{(0)}}
\left[ \delta \phi~ +~ x^+ \partial_+ \delta \phi~ +~ x^-
\partial_- \delta \phi \right] \right) \cr j_-~~ &=~~ 2 \lambda
\partial_- \left( e^{-2\phi^{(0)}} \left[ \delta \phi~ +~ x^+
\partial_+ \delta \phi~ +~ x^- \partial_- \delta \phi \right]
\right). \cr} \eqno(2.11)$$
Conservation of this current, $\nabla \cdot j = 0$, implies the
existence of two charges,
$$\eqalign{Q^+~~ &=~~ \int^{\scri_L^+} dx^- j_- \cr Q^-~~ &=~~
\int^{\scri_R^+} dx^+ j_+ \cr} \eqno(2.12)$$
which are constant along $x^+$ and $x^-$ respectively. The current
densities are total derivatives and their integrals can be measured
as surface terms on $\scri^+$. For example, 
$$Q^-~~ =~~ 2 \lambda \left( e^{-2\phi^{(0)}} \left[ \delta \phi~
+~ x^+ \partial_+ \delta \phi~ +~ x^- \partial_- \delta \phi
\right] \right)_{{\cal I}_R^+}~~ =~~ +~ M~~ >~~ 0. \eqno(2.13)$$
is the Bondi mass on $\scri_R^+$ and similarly for $\scri_L^+$.

To introduce dynamical matter fields into the problem, consider a
shock wave of incoming matter travelling in the $x^-$ direction
with strength $a$ at constant $x^+ = x_0^+$
$${1 \over 2}~ \partial_+ f~ \partial_+ f~~ =~~ a \delta(x^+~ -~
x^+_0). \eqno(2.14)$$
The equations of motion are solved by
$$\sigma~~ =~~ \lambda^2 x^+ x^-~ -~ a (x^+~ -~ x^+_0) \Theta (x^+~
-~ x^+_0), \eqno(2.15)$$
which is the linear dilaton vacuum when $x^+ < x^+_0$, and the
naked singularity with mass $M = ax^+_0\lambda$ when $x^+ > x^+_0$. 
The resulting metric and curvature are identical to those in (2.6)
and (2.7), with a shift in retarded time, $x^- \rightarrow x^- -
a/\lambda^2$.  The mass of the singularity on $\scri_R^+$ is now
$a\lambda x_0^+$ and the Penrose diagram for the spacetime is given
in figure II. We should note that this solution does not actually
describe the formation of a naked singularity as the latter appears
together with the incoming shock wave and not in its future,
contrasting with the scenario for the formation of a black hole. 

As in any two dimensional system, Wald's axioms and the trace
anomaly can be used to study semi-classical quantum effects in this
model. In two dimensions the conservation equations 
$$\nabla_\mu \langle T_{\mu\nu} \rangle~~ =~~ 0 \eqno(2.16)$$
and the trace anomaly 
$$\langle T^\mu_\mu \rangle~~ =~~ -~\alpha R\eqno(2.17)$$
together completely determine the stress tensor of the matter
fields up to a boundary condition dependent term. The trace
anomaly, (2.17), gives the one loop correction to the stress energy
tensor and $\alpha$ is a positive spin dependent constant (it is
$1/24\pi$ for scalar fields). Integrating (2.16), using (2.17) then
gives 
$$\eqalign{\langle T_{++}\rangle~~ &=~~ -~ \int~ {{dx^-} \over
\sigma} \partial_+ (\sigma \langle T_{+-}\rangle)~~ +~~ A(x^+) \cr
\langle T_{--}\rangle~~ &=~~ -~ \int~ {{dx^+} \over \sigma}
\partial_- (\sigma \langle T_{+-}\rangle)~~ +~~ B(x^-) \cr}
\eqno(2.18)$$
where $A(x^+)$ and $B(x^-)$ are the boundary condition dependent
terms mentioned before and which we fix by the following physical
requirements: (a) there should be no incoming radiation and (b) the
stress tensor should vanish exactly in the linear dilaton vacuum
(for $x^+ < x_0^+$). These conditions imply that 
$$A(x^+)~~ =~~ {1 \over {2x^{+2}}},~~~~~~~~~~~~ B(x^-)~~ =~~ {1
\over {2x^{-2}}} \eqno(2.19)$$
In a coordinate system in which the metric is asymptotically flat,
defined by
$$\eqalign{x^+~~ &=~~ {1 \over \lambda}~ e^{\lambda \sigma^+} \cr
x^-~~ &=~~ {1 \over \lambda}~ e^{\lambda \sigma^-}~ +~ {a \over
{\lambda^2}}. \cr} \eqno(2.20)$$
we find that $\langle T_{\mu\nu} \rangle$ has the following
behavior
$$\eqalign{\langle T^{(\sigma)}_{++}\rangle~~ & \rightarrow~~
0,~~~~~~~~~~ \langle T^{(\sigma)}_{+-} \rangle~~ \rightarrow~~ 0
\cr \langle T^{(\sigma)}_{--}\rangle~~ &=~~  {{\alpha \lambda^2}
\over 2}~ \left[ 1~~ -~~ {1 \over {(1~ +~ (a/\lambda) e^{-\lambda
\sigma^-})^2}} \right]. \cr} \eqno(2.21)$$
on $\scri_R^+$ and all components of the tensor vanish on
$\scri_L^+$. 

$\langle T^{(\sigma)}_{--} \rangle$ is the flux across $\scri_R^+$
and its behavior is remarkably different from that calculated by
CGHS for the black hole. It approaches a maximum, $\alpha
\lambda^2/2$ in the far past of $\scri_R^+$ as $x^- \rightarrow
a/\lambda^2$ and decreases smoothly to zero as $i^+$ is approached,
i.e., as $x^- \rightarrow \infty$.  The total energy radiated is,
of course, the integrated flux, $\int_{-\infty}^{\sigma^-} d
\sigma^- \langle T^{(\sigma)}_{--} \rangle$, but this is infinite
because the flux approaches a steady state in the infinite past, as
opposed to black hole evaporation in the same model where the flux
is shown to increase steadily from zero on $\scri_R^-$ to a steady
state independent of the mass of the hole as the horizon of the
hole is approached. Infinite energy cannot be radiated from a
system with a finite amount of energy and the paradox expresses the
fact that, beyond a certain point, one must account for the back
reaction of the spacetime. It is safe, however, to say that this is
at least an indication that naked singularities, if they form, tend
to radiate cataclysmically (``explode''), as opposed to black holes
that prefer to radiate slowly or ``evaporate.''

The back reaction of the spacetime becomes important when the
dilaton coupling gets large, that is the size of the dilaton
coupling indicates the extent to which the effects of the spacetime
itself on the matter system should be taken into account. However,
the value of the dilaton field at the point at which the
singularity bursts into radiation
$$e^\phi~~ =~~ {1 \over {\sqrt{ax_0^+}}} \eqno(2.22)$$
depends on the inverse square root of its Bondi mass and
consequently is small when the classical singularity is very
massive. The one loop approximation above, ignoring the back
reaction, therefore seems valid in this limit. However, as the mass
decreases due to the evaporation, the dilaton coupling will grow
and the above expressions will break down.
\vskip 0.25in

{\noindent \bf III. Consistent Potentials}

It would be nice to see if the general features of the Hawking
radiation from both spacelike and timelike singularities are
reproduced in other models of two dimensional gravity. These models
must be consistent generalizations of the CGHS model, in the sense
that consistency means maintaining scale and diffeomorphism
invariance as explained in the introduction. This leads to a
consideration of what form the full quantum theory might take.

To incorporate quantum effects to lowest order in studying the back
reaction, one must include the contribution of the conformal
anomaly (which arises because the measure on the space of matter
fields is non-invariant) in the effective action as we have done
above.  However, a consistent quantization of the theory in the
conformal gauge requires that both the kinetic and potential terms
get renormalized in such a manner that the theory becomes a
conformal field theory.  The requirements of scale and
diffeomorphism invariance have been checked for the CGHS model by
de Alwis and by Bilal and Calan.${}^7$ If $f$-field potentials are
included as in (1.1), their analysis must be modified and will
yield a restriction on the allowed functionals $U(\vf)$.  Below we
derive the condition that makes for a consistent theory. Our
arguments will closely follow those of de Alwis in ref.[7]

In general terms, we can expect that the full quantum action,
including matter but not including ghosts takes the form (in terms
of some fiducial metric $g = e^{2\rho} \hg$)
$$S[X,\hg]~~ =~~ \int d^2x {\sqrt {\hg}} \left[ - {1 \over 2}
\hg^{\mu\nu} G_{ab} \nabla_\mu X^a \nabla_\nu X^b~ -~ \hR \Phi~ +~
T(X) \right] \eqno(3.1)$$
where $T$ is the tachyon potential and $X^a$ is the $N+2$
dimensional vector
$$X^a~~ =~~ \left[ \matrix{\phi \cr \rho \cr f_1 \cr . \cr . \cr .
\cr f_N} \right] \eqno(3.2)$$
including the $N$ matter fields. The action in (1.1) is to be
viewed as the weak coupling limit of (3.1), i.e., in the limit
$e^{2\phi} << 1$ and including the anomaly term, (3.1) should look
like
$$\eqalign{S[\phi,\rho,f_i]~~ &=~~ \int d^2x {\sqrt{\hg}} \left[
e^{-2\phi} \left( 4(\hn \phi)^2~ -~ 4 \hn \phi \cdot \hn \rho
\right)~ -~ \alpha (\hn \rho)^2 \right. \cr &\left. -~\hR \left(
e^{-2\phi}~ -~ \alpha \rho\right)~ +~ \Lambda e^{2(\rho-
\phi)}\right.\cr &\left. -~ {1 \over 2} \sum_i (\hn f_i)^2~ +~ 4
\mu^2 e^{2(\rho-\phi)}U(\vf)\right]\cr} \eqno(3.3)$$
Comparing (3.1) and (3.3), we find that the low energy limit is
given by
$$\eqalign{G_{ab}~~ &=~~ \left[\matrix{-8e^{-2\phi} & 4e^{-2\phi}
& 0 & . & . & . & 0 \cr 4e^{-2\phi} & 2\alpha & 0 & . & . & . & 0
\cr 0 & 0 & 1 & . & . & . & . \cr 0 & 0 & 0 & 1 & . & . & . \cr .
& . & . & . & . & . & .\cr. & . & . & . & . & . & 1} \right] \cr
\Phi~~ &=~~ +~ \alpha \rho~ -~ e^{-2\phi} \cr T(X)~~ &=~~ \Lambda
e^{2(\rho-\phi)}~ +~ 4 \mu^2 e^{2(\rho-\phi)} U(\vf). \cr}
\eqno(3.4)$$
Now consider a generalization of the above field space metric, but
with the appropriate limit
$$ds^2~ =~ -8e^{-2\phi} (1 + h_1(\phi)) d\phi^2~ +~ 8e^{-2\phi} (1
+ h_2(\phi)) d\phi d\rho~ +~ 2\alpha (1 + h_3(\phi)) d\rho^2~ +~
\sum_i df_i^2 \eqno(3.5)$$
where $h_i(\phi)$ are $O(e^{2\phi})$ functionals of the dilaton,
$\phi$. We will restrict ourselves to the case $h_3(\phi)=0$, which
leads to an exactly solvable model. Then the metric can be brought
to a flat form by the following field redefinitions:
$$\eqalign{\Omega~~ &=~~ {\sqrt{2|\alpha|}} \left[ \rho~ -~ {1
\over \alpha} e^{-2\phi}~ +~ {2 \over \alpha} \int e^{-2\phi}
h_2(\phi) d\phi \right]\cr \chi~~ &=~~ \int P(\phi)
d\phi\cr}\eqno(3.6)$$
where
$$P(\phi)~~ =~~ {{e^{-2\phi}} \over {\sqrt{|\alpha|}}} \left[8
\alpha e^{2\phi} (1 + h_1)~ +~ 8 (1 + h_2)^2 \right]^{1 \over 2}
\eqno(3.7)$$
which give for the metric in (3.5)
$$ds^2~~ =~~ \mp d\chi^2~~ \pm~~ d\Omega^2~~ +~~ \sum_i df_i^2.
\eqno(3.8)$$
where the upper sign is to be used for $\alpha > 0$ and the lower
sign for $\alpha < 0$. In the weak field limit, 
$$\eqalign{\Omega~~ &\sim~~ {\sqrt{2|\alpha|}} \left( \rho~ -~ {1
\over \alpha} e^{-2\phi}~ +~ . . . \right) \cr \chi~~ &\sim~~ 2
{\sqrt{2 \over {|\alpha|}}} \left( {\alpha \over 2} \phi ~ -~ {1
\over 2} e^{-2\phi}~ +~ ... \right)\cr} \eqno(3.9)$$
Let us now consider the beta-function equations for the action in
(3.1)
$$\eqalign{0~~ =~~ \beta^G_{ab}~~ &=~~ - {\cR}_{ab}~ +~ 2 \nabla_a
\nabla_b \Phi~ -~ \nabla_a T \nabla_b T~ +~ .... \cr 0~~ =~~
\beta^\Phi~~ &=~~ - {\cR}~ +~ 4 G^{ab} \nabla_a \Phi \nabla_b \Phi~ 
-~ 4 \nabla^2 \Phi~ +~ {{N-24} \over 3}~ +~ . . . . \cr 0~~ =~~ 
\beta^T~~ &=~~ -2 \nabla^2 T~ +~ 4 G^{ab} \nabla_a \Phi \nabla_b T~
-~ 4T~ +~ . . . \cr} \eqno(3.10)$$
It is easy to see, using (3.8), that the first of the above
equations implies that $\Phi$ is a linear function of the fields,
and to have the appropriate limit it must be uniquely given by
$$\Phi~~ =~~ \pm {\sqrt{{|\alpha|} \over 2}} \Omega \eqno(3.11)$$
The second beta function equation gives trivially $\alpha = (24-
N)/6$, and the third implies that
$$\pm \partial_\chi^2 T~ \mp~ \partial_\Omega^2 T~ \pm~
{\sqrt{2|\alpha|}} \partial_\Omega T~ -~ \sum_i \partial_{f_i}^2 T~
-~ 2T~~ =~~ 0 \eqno(3.12)$$
where we have retained only first order terms in the tachyon
potential. Consider, first, that $T(X)$ depends only on the fields
$(\chi,\Omega)$. Obviously, a solution that will reduce to the
appropriate one in the limit of weak coupling is
$$T(\chi,\Omega)~~ =~~ \Gamma e^{{\sqrt{2\over {|\alpha|}}} (\Omega
\mp \chi)}~~ \sim~~ \Gamma e^{2(\rho-\phi)} \eqno(3.13)$$
where $\Gamma$ is some constant. 

Including the $f$ fields, we can now obtain a condition for what
constitutes a consistent potential, coupled as in (1.1) to the
metric. Thus, if $T$ is of the form 
$$T(X)~~ \sim~~ e^{{\sqrt{2\over |\alpha|}} (\Omega \mp \chi)}
U(\vf) \eqno(3.14)$$
i.e., if $T(\chi,\Omega,f_i)$ has the same coupling to the $\rho$
and $\phi$ fields as the cosmological term, then (3.12) can be
satisfied only if $U(\vf)$ is harmonic,
$$\sum_i \partial^2_{f_i} U(\vf)~~ =~~ 0 \eqno(3.15)$$
The general solution can therefore be expressed as arbitrary sums
over functions of the form
$$U_k(\vf)~~ =~~ e^{i{\vec \omega}_k \cdot \vf} \eqno(3.16)$$
where the ${\vec \omega}_k$ satisfy $\omega_k^2 = 0$. For example, 
if one restricts the number, $N$, of fields, $f_i$, to be even, the
potential 
$$U(\vf)~~ =~~ \prod_{i=1}^{N/2} \cos f_i \prod_{j=N/2+1}^N \cosh
f_j \eqno(3.17)$$
satisfies the condition (3.15). Now (3.12) is linear in $T$ (to
this approximation), so
$$T(X)~~ =~~ \Lambda e^{2(\rho-\phi)}~~ +~~ 4 \mu^2 e^{2(\rho -
\phi)} (\prod_{i=1}^{N/2} \cos f_i \prod_{j=N/2+1}^N \cosh f_j~ -~
1) \eqno(3.18)$$
is a consistent potential.
\vskip 0.25in

{\noindent \bf IV. Singularities from sine-Gordon Solitons}

Before going on to examine (3.1) in detail, we note that the sine-
Gordon theory is a special case of (3.18).  This follows directly
by taking $f_i = 0$ for $i \geq 2$.  Therefore, here we will
examine the singularities induced by sine-Gordon solitons in the
Witten model. They are qualitatively different from the singularity
discussed in section II being neither purely timelike nor purely
spacelike but a combination of the two. Thus they combine both
black holes and naked singularities, but the Hawking evaporation of
the solitons will depend only on their asymptotic properties.
Consider the action in (1.1) with the sine-Gordon potential,
$$U(f)~~ =~~ \left( \cos f~ -~ 1 \right), \eqno(4.1)$$
setting $f=f_1 \neq 0$ and $f_i=0$ for all $i \geq 2$ in (3.18). As
in section II, we will treat the action with a positive
cosmological constant, $\Lambda = +4\lambda^2$, and with a negative
cosmological constant, $\Lambda = -4\lambda^2$, separately. When
$\Lambda = 4 \lambda^2$, the classical constraints and equations of
motion can be written in light cone coordinates as follows
$$\eqalign{0~~ &=~~ {\cal T}_{++}~~ =~~ e^{-2\phi} \left[ -4
\partial_+ \rho \partial_+ \phi~ +~ 2 \partial_+^2 \phi \right]~ -~
{1 \over 2} (\partial_+ f)^2\cr0~~ &=~~ {\cal T}_{--}~~ =~~ e^{-
2\phi} \left[-4 \partial_- \rho \partial_- \phi~ +~ 2 \partial_-^2
\phi \right]~ -~ {1 \over 2} (\partial_- f)^2\cr 0~~ &=~~ {\cal
T}_{+-}~~ =~~ e^{-2\phi} \left[-2 \partial_+ \partial_- \phi~ +~ 4
\partial _+ \phi \partial_- \phi~ +~ {\Lambda \over 4} e^{2 \rho}~
+~ \mu^2 e^{2 \rho} (\cos f~ -~ 1) \right]\cr &-4 \partial_+
\partial_- \phi~ +~ 4 \partial_+ \phi \partial_- \phi~ +~ 2
\partial_+ \partial_- \rho~ +~e^{2 \rho} \left[{\Lambda \over 4}~
+~ \mu^2 (\cos f~ -~ 1) \right]~~ =~~ 0\cr & +\partial_+ \partial_-
f~~ +~~\mu^2 e^{2(\rho-\phi)} \sin f~~ =~~ 0\cr} \eqno(4.2)$$
with the solution in conformal gauge  
$$\eqalign{f_{kink}~~ &=~~ 4 \tan^{-1} e^{(\Delta~ -~ \Delta_0)}\cr
\sigma~~ &=~~ a~ +~ bx^+~ +~ cx^-~ -~ {\Lambda \over 4} x^+ x^-~ -~
2 \ln \cosh(\Delta~ -~ \Delta_0)\cr} \eqno(4.3)$$
in terms of $\Delta = \gamma_+ x^+~ +~ \gamma_- x^-$, where
$$\gamma_\pm~~ =~~ \pm~ \mu \sqrt{{1 \pm v} \over {1 \mp v}},
\eqno(4.4)$$ 
$v$ is the velocity of the soliton, $f(x,t) = f(x+vt)$, $\Delta =
\Delta_0$ is its center which we take without loss of generality to
be greater than or equal to zero, and $a$, $b$ and $c$ are
arbitrary constants.  There is also the antikink solution 
$$\eqalign{f_{antikink}~~ &=~~ 4 \arctan e^{-(\Delta~ -~ \Delta_0)}
\cr \sigma~~ &=~~ -~ {\Lambda \over 4} x^+ x^-~ -~ 2 \ln \cosh
(\Delta~ -~ \Delta_0)\cr} \eqno(4.5)$$
which may be analyzed along the same lines. To fix the constants
$a$, $b$ and $c$ we require that the metric reduces to the linear
dilaton vacuum in the absence of the incoming soliton, i.e., in the
limit as the soliton stress energy, $T^f_{\mu\nu} \rightarrow 0$,
$\Delta_0 \rightarrow 0$. This gives for the solution in (4.3)
$$\sigma~~ =~~ -~ {\Lambda \over 4} x^+ x^-~ -~ 2 \ln \cosh(\Delta~
-~ \Delta_0). \eqno(4.6)$$
The curvature singularity is at $\sigma(x) = 0$.  When $\Lambda >
0$ (4.5) represents a positive energy singularity combining a white
hole a timelike singularity and a black hole, all smoothly joined
along the soliton center (by a white hole we mean a spacelike naked
singularity). On the contrary, when $\Lambda < 0$ it represents two
naked singularities smoothly joined at the soliton center. For
positive cosmological constant, using the (timelike) Killing vector
$\xi^\mu = (x^{+'}, x^{-'})$ where
$$\eqalign{x^{+'}~~ &=~~ x^+~ +~ {{2 \gamma_-} \over
{\lambda^2}}\cr x^{-'}~~ &=~~ x^-~ +~ {{2 \gamma_+} \over
{\lambda^2}}\cr}\eqno(4.7)$$
and a linearization of ${\cal T}_{\mu\nu}$ about the dilaton
vacuum, the conserved charge or the Bondi mass of the singularity
is found to be 
$$M_R~~ =~~ 2\lambda \left(\ln 2~ -~ {{2\mu^2} \over {\lambda^2}}~
+~ \Delta_0 \right).\eqno(4.8)$$
on $\scri_R^+$ and 
$$M_L~~ =~~ 2\lambda \left(\ln 2~ -~ {{2\mu^2} \over {\lambda^2}}~
-~ \Delta_0 \right)\eqno(4.9)$$
on $\scri_L^+$. Without any loss of generality we will assume that
$\Delta_0 > 0$ throughout, in which case the soliton never actually
enters the left region. The Kruskal diagram is displayed in figure
III. The soliton is seen to emerge from the merging of a white
hole, and a timelike singularity at $(x^+ = 0,~ x^- =
\Delta_0/\gamma_-)$, the white hole extending from $(x^+ = 2
\gamma_-/\lambda^2,~ x^- = - \infty)$ on $\scri_R^-$ to $(x^+ = 0,~
x^- = \Delta_0/ \gamma_-)$ where it smoothly turns into a timelike
line proceeding to $x^- = 0,~ x^+ = \Delta_0/\gamma_+)$ at which
point the soliton is reabsorbed. Here the singularity once again
turns spacelike smoothly and reaches $\scri_R^+$ at $(x^+ = \infty,
x^- = - 2\gamma_+/\lambda^2)$. The singularities are asymptotically
spacelike and the metric approaches
$$\sigma~~ \rightarrow~~ -~ \lambda^2 x^{+'} x^{-'}~~ +~~
2\Delta_0~ -~ {{4 \mu^2} \over {\lambda^2}}~ +~ 2 \ln 2
\eqno(4.10)$$
near $\scri_R^+$. Although the singularities in figure III have
been drawn for a specific value of the parameters, their
qualitative behavior does not differ if both the left and right
Bondi masses are positive. 

As $\Delta_0$ approaches zero (figure IV) the left and right
singularities merge forming, in the limit, a white hole that
extends from $x^+ = -2\gamma_-/\lambda^2$ on ${\cal I}_R^-$ to $x^-
= 2\gamma_+ / \lambda^2$ on ${\cal I}_L^-$ and a black hole that
stretches from $x^- = -2\gamma_+ /\lambda^2$ on ${\cal I}_R^+$ to
$x^+ = 2\gamma_- / \lambda^2$ on ${\cal I}_L^+$, intersecting at
the origin. The length of the timelike naked singularity has shrunk
to zero and the masses measured on both null infinities are now the
same. Even if they are zero ($2 \mu^2 = \lambda^2 \ln 2 $) soliton
energy and momentum is present throughout the spacetime. 

The spacetime with negative cosmological constant ($\Lambda = - 4
\lambda^2$) 
$$\sigma~~ =~~ \lambda^2 x^+ x^-~ -~ 2 \ln \cosh(\Delta~ -~
\Delta_0) \eqno(4.11)$$
is shown in figure V. All singularities are timelike as they
approach ${\cal I}^\pm$. In the top region the timelike singularity
approaches ${\cal I}_R^+$ at ($x^+ = \infty, x^- = 2\gamma_+/
\lambda^2)$ and approaches ${\cal I}_L^+$ at $(x^+ = - 2 \gamma_-
/ \lambda^2, x^- = \infty)$. The two timelike sections merge at a
white hole in the region where the soliton center enters the
spacetime. The latter emerges at $(x^+ = \Delta_0 / \gamma_+, x^-
= 0)$ and travels to $i^0$. The Bondi mass depends on whether the
asymptotic observer is located on ${\cal I}_R^+$ or ${\cal I}_L^+$,
$$\eqalign{M_R~~ &=~~ 2\lambda \left({{2\mu^2} \over {\lambda^2}}~
+~ \Delta_0~ +~ \ln 2 \right)\cr {\rm and}~~~~~ M_L~~ &=~~ 2\lambda
\left({{2\mu^2} \over {\lambda^2}}~ -~ \Delta_0~ +~ \ln 2
\right),\cr} \eqno(4.12)$$
the qualitative behavior again being independent of the parameter
values chosen if both masses are positive. The masses are equal
when $\Delta_0 = 0$. In this limit one has two naked singularities,
the first extending from $x^- = - 2 \gamma_+/ \lambda^2$ on ${\cal
I}_L^-$ to $x^- = 2 \gamma_+/ \lambda^2$ on ${\cal I}_R^+$ and the
other from $x^+ = 2 \gamma_-/ \lambda^2$ on ${\cal I}_R^-$ to $x^+
= - 2 \gamma_-/ \lambda^2$ on ${\cal I}_L^+$, intersecting at the
origin (figure VI). 

Though our emphasis here is on the quantum behavior of
(asymptotically) timelike singularities, this model also provides
asymptotically spacelike singularities and it is instructive to
compare the quantum behaviors of the two. We shall therefore
examine the Hawking radiation both for positive and negative
cosmological constant. For positive cosmological constant, the
classical stress tensor of the $f-$ soliton (right hand quadrant of
figure III) is exponentially vanishing on $\scri^+$ so that Hawking
radiation is the dominant effect there. Again, the only possible
one loop contribution to the trace of the stress energy tensor must
be of the form $-\alpha R$, $R$ being the only available geometric
invariant. Thus we expect the quantum correction to take the form
$${T^{(q)\mu}}_\mu~~ =~~ -~ 4 \sigma T^q_{+-}~~ =~~ -~ \alpha R~~
=~~ -~ 4 \alpha \sigma \partial_+ \partial_- \ln \sigma
\eqno(4.13)$$
for some (positive) dimensionless constant $\alpha$. The two
conservation equations can once again be integrated and the full
stress tensor expressed in terms of two arbitrary functions
$A(x^+)$ and $B(x^-)$ as before in (2.18). A consistent solution
should admit no incoming radiation on ${\cal I}_R^-$ other than 
any matter fields that might be present and vanish in the absence
of the soliton, that is, in the linear dilaton vacuum.  The stress
tensor satisfying these conditions is
$$\eqalign{\langle T_{++}\rangle~~ &=~~ T^f_{++}~ +~ T^q_{++}~~ =~~
T^f_{++}~~ -~~ \alpha \left( {{\partial^2_+ \sigma} \over \sigma}~
-~ {1 \over 2} \left[{{\partial_+ \sigma} \over \sigma}\right]^2
\right)~ -~  {\alpha \over {2 x^{+'2}}}\cr \langle T_{--}\rangle~~
&=~~ T^f_{--}~ +~ T^q_{--}~~ =~~ T^f_{--}~~ -~~ \alpha \left(
{{\partial^2_- \sigma} \over \sigma}~ -~ {1 \over 2}
\left[{{\partial_- \sigma} \over \sigma}\right]^2 \right)~ -~ 
{\alpha \over {2 x^{-2}}}\cr \langle T_{+-}\rangle ~~ &=~~ T^f_{+-
}~~ +~~ \alpha \partial_+ \partial_- \ln \sigma\cr} \eqno(4.14)$$
where $x^{+'} = x^+ + 2 \gamma_-/\lambda^2$. It is most convenient
to analyze the above expressions in the coordinate system in which
the metric is manifestly asymptotically flat. Define, therefore the
coordinates $\sigma^\pm = t \pm x$ by
$$\eqalign{x^+~~ =~~ {1 \over \lambda} e^{\lambda \sigma^+}~ -~ {{2
\gamma_-} \over {\lambda^2}}\cr x^-~~ =~~ -~ {1 \over \lambda} e^{-
\lambda \sigma^-}~ -~ {{2 \gamma_+} \over {\lambda^2}}\cr}
\eqno(4.15)$$
Thus $\sigma^+ \rightarrow \infty$ corresponds to the lightlike
surface $x^+ = \infty$ and $\sigma^- \rightarrow \infty$ to the
lightlike surface $x^- = - \infty$, while $\sigma^- \rightarrow
\infty$ corresponds to the lightlike surface $x^- = - 2 \gamma_- /
\lambda^2$. Transforming the expressions in (4.14) to the new
system, we find that 
$$\langle T^{(\sigma)}_{++}\rangle ~ \rightarrow~ 0,~~~~~~~~
\langle T^{(\sigma)}_{+-}\rangle ~ \rightarrow~ 0 \eqno(4.16)$$
and
$$\langle T^{(\sigma)}_{--}\rangle~~ \rightarrow~~ {{\alpha
\lambda^2} \over 2} \left [ 1~ -~ {1 \over {\left( 1 + {{2
\gamma_+} \over \lambda} e^{\lambda \sigma^-} \right )^2 }}\right].
\eqno(4.17)$$
Notice that the form of the solution is precisely the same as the
solution for the collapsing shock wave. $\langle T^{(\sigma)}_{--
}\rangle$ is the outgoing flux at ${\cal I}_R^+$. It grows smoothly
from zero at $x^- = - \infty$ to a maximum of $\alpha \lambda^2 /2$
at $x^- = - 2 \gamma_+ / \lambda^2$ on ${\cal I}_R^+$ and depends
on the soliton mass parameter but not on the mass of the
singularity itself. This would seem to be a general feature of the
Hawking evaporation from asymptotically spacelike singularities in
these two dimensional models with collapsing matter, having been
shown to be true for the radiation from a black hole formed by an
incoming shock wave in the CGHS model. The integrated flux along
${\cal I}_R^+$ is the total energy lost by the incoming soliton,
but as the flux rapidly approaches its maximum value of $\alpha
\lambda^2 / 2$, the integrated flux shows an infinite loss of
energy if the integral is performed up to the future horizon at
$x^- = - 2\gamma_+ / \lambda^2$. However, as CGHS pointed out, this
is a consequence of having neglected the back reaction and must not
be taken seriously. Instead, one can try to estimate the retarded
time, $x^-_\tau$ at which the integrated Hawking radiation is equal
to the mass of the singularity, $M = 4 \lambda (\Delta_0 - 2 \mu^2/
\lambda^2 + \ln 2)$. We find
$$\int_{-\infty}^{\sigma^-_\tau} d\sigma^- \langle T^{(\sigma)}_{--
} \rangle~~ =~~ {{\alpha\lambda} \over 2} \left[ 1~ -~ {1 \over
{\left( 1 + {{2 \gamma_+} \over \lambda} e^{\lambda \sigma^-_\tau}
\right )}}~ +~ \ln \left(1 + {{2 \gamma_+} \over \lambda}
e^{\lambda \sigma^-_\tau} \right)\right]~~ =~~ M \eqno(4.18)$$
For a small mass singularity, the retarded time is given by
$$x^-_\tau~~ =~~ -~ {{2 \gamma_+} \over {\lambda^2}}\left( 1~ +~
{{\alpha \lambda} \over M}\right) \eqno(4.19)$$
which, when na\"\i vely traced backwards, corresponds to the point
$$(x^+_\tau, x^-_\tau)~~ =~~ \left( {{\Delta_0} \over {\gamma_+}}~
+~ {{2 \gamma_-} \over {\lambda^2}} \left[ 1~ +~ {{\alpha \lambda}
\over M} \right],- {{2 \gamma_+} \over {\lambda^2}} \left[ 1~ +~
{{\alpha \lambda} \over M}\right] \right) \eqno(4.20)$$
on the soliton trajectory. If $x^+_\tau < 0$, the soliton energy
has evaporated earlier than the appearance of its center in the
spacetime. An observer on ${\cal I}_R^+$ sees a white hole which
rapidly radiates away all its energy.  This of course is true only
if the mass of the singularity is small. On the other hand, if
$x^+_\tau$ is greater than zero, the soliton does enter the
spacetime evaporating eventually by $x^-_\tau$ in (4.19). How
reliable is the estimate of its lifetime above? Assuming that the
soliton center does enter the spacetime before evaporating
completely, the dilaton coupling constant at the turn around point,
$$e^\phi~~ =~~ {1 \over {\sqrt{2 \left(1~ +~ {{\alpha \lambda}
\over M}\right) \left[ {M \over {4\lambda}}~ -~ \ln 2~ -~ {{2\mu^2
\alpha} \over{\lambda M}} \right]}}} \eqno(4.21)$$
is large for a small mass singularity and signals the breakdown of
the one loop approximation if the back reaction is ignored. 

On the other hand, if $M$ is large, 
$$\int_{-\infty}^{\sigma^-_\tau} d\sigma^- \langle T^{(\sigma)}_{--
} \rangle~~ \sim~~ {{\alpha \lambda} \over 2}~ \left[ \ln
{{2\gamma_+} \over \lambda}~ +~ \lambda \sigma^-_\tau \right]~~ =~~
M\eqno(4.22)$$
or
$$x^-_\tau~~ =~~ -~ {{2\gamma_+} \over {\lambda^2}} \left[1~ +~
e^{-2M/\alpha \lambda} \right] \eqno(4.23)$$
which, when traced back corresponds to the point
$$(x^+_\tau, x^-_\tau)~~ =~~ \left({{\Delta_0} \over {\gamma_+}}~
+~ {{2 \gamma_-} \over {\lambda^2}} \left[ 1~ +~ e^{-
2M/\alpha\lambda} \right], -~ {{2 \gamma_+} \over {\lambda^2}}
\left[ 1~ +~ e^{-2M/\alpha\lambda} \right]\right)\eqno(4.24)$$
on the soliton trajectory. The dilaton coupling at this point has
the value
$$e^{\phi}~~ =~~ {1 \over {\sqrt{2 \left( 1~ +~ e^{-2M/\alpha
\lambda} \right) \left( {M \over {4 \lambda}}~ -~ \ln 2~ -~ 
{{2\mu^2} \over {\lambda}} e^{-2M/\alpha \lambda} \right)}}}
\eqno(4.25)$$
and is small in the limit of large $M$. The soliton evaporates
completely by the time the observer has reached the event horizon
and the black hole never forms. Moreover this is the limit in which
the one loop approximation is a satisfactory indication of what may
actually be happening.  Similar conclusions can be drawn in the
CGHS (shock wave) model.  In the low mass limit that the dilaton
coupling is large at the turn around point, being proportional only
to $\sqrt{1/\alpha}$. However, in the large mass limit the dilaton
coupling behaves as the inverse square root of the mass. Thus,
there seems to be no essentially new feature in the evaporation of
the soliton.

Next consider an observer in the left quadrant of figure III. As we
have mentioned, because of our choice of $\Delta _0 > 0$ the
soliton center never enters this region and this observer lives in
a universe inhabited only by the tail of the soliton energy and the
singularity described earlier. The appropriate boundary conditions
on the Hawking stress tensor are (a) its vanishing in the absence
of the soliton and (b) no flux across ${\cal I}_L^-$.  It follows
that the quantum contribution to the stress tensor is given by
$$\eqalign{T^q_{++}~~ &=~~ -~~ \alpha \left( {{\partial^2_+ \sigma}
\over \sigma}~ -~ {1 \over 2} \left[{{\partial_+ \sigma} \over
\sigma}\right]^2 \right)~ -~  {\alpha \over {2 x^{+2}}}\cr T^q_{--
}~~ &=~~ -~ \alpha \left( {{\partial^2_- \sigma} \over \sigma}~ -~
{1 \over 2} \left[{{\partial_- \sigma} \over \sigma}\right]^2
\right)~ -~  {\alpha \over {2 x^{-'2}}}\cr T^q_{+-}~~ &=~~ \alpha
\partial_+ \partial_- \ln \sigma \cr} \eqno(4.26)$$
where $x^{-'} = x^-~ -~ 2 \gamma_+ /\lambda^2$. Going to the system
$\sigma^\pm = t \pm x$ given by
$$\eqalign{x^+~~ &=~~ -~ {1 \over \lambda} e^{- \lambda \sigma^+}~
+~ {{2 \gamma_-} \over {\lambda^2}}\cr x^-~~ &=~~ {1 \over \lambda}
e^{\lambda \sigma^-}~ +~ {{2 \gamma_+} \over {\lambda^2}}\cr}
\eqno(4.27)$$
in which the metric is manifestly flat at null infinity, ($\sigma^+
\rightarrow \infty$ corresponds to the lightlike line $x^+ = 2
\gamma_-/\lambda^2$ and $\sigma^- \rightarrow -  \infty$ to the
lightlike line $x^+ = 2 \gamma_+ / \lambda^2$ while $\sigma^+
\rightarrow - \infty$ and $\sigma^- \rightarrow \infty$ correspond
to the respective lightlike infinities) we obtain (on ${\cal
I}_L^+$)
$$\langle T^{(\sigma)}_{--}\rangle ~ \rightarrow~ 0,~~~~~~~~
\langle T^{(\sigma)}_{+-}\rangle ~ \rightarrow~ 0 \eqno(4.28)$$
and
$$\langle T^{(\sigma)}_{++}\rangle~~ =~~ {{\alpha \lambda^2} \over
2} \left [ 1~ -~ {1 \over {\left( 1 - {{2 \gamma_-} \over \lambda}
e^{\lambda \sigma^+} \right )^2 }}\right] \eqno(4.29)$$
In this part of the world, the radiation also grows steadily from
zero in the infinite past to its maximum value (of $\alpha
\lambda^2 /2$) in the far future of ${\cal I}_L^+$. The integrated
flux is infinite and, as we have argued before, that is a signal
that the back reaction is to be taken into account.

We turn now to solitons in 2d gravity with a negative cosmological
constant. We consider the observer in the top quadrangle. In the
absence of ${\cal I}^-$ the only reasonable boundary condition one
may impose upon the stress tensor is that it vanishes in the
absence of the soliton, a limit we have defined earlier. The
quantum corrections to the stress tensor then take the form
$$\eqalign{T^q_{++}~~ &=~~ -~~ \alpha \left( {{\partial^2_+ \sigma}
\over \sigma}~ -~ {1 \over 2} \left[{{\partial_+ \sigma} \over
\sigma}\right]^2 \right)~ -~  {\alpha \over {2 x^{+2}}}\cr T^q_{--
}~~ &=~~ -~ \alpha \left( {{\partial^2_- \sigma} \over \sigma}~ -~
{1 \over 2} \left[{{\partial_- \sigma} \over \sigma}\right]^2
\right)~ -~  {\alpha \over {2 x^{-2}}}\cr T^q_{+-}~~ &=~~ \alpha
\partial_+ \partial_- \ln \sigma \cr} \eqno(4.30)$$
Again, to analyze the tensors it is convenient to go to a system in
which the metric is asymptotically flat: define the system
$\sigma^\pm = t \pm x$ by
$$\eqalign{x^+~~ &=~~ {1 \over \lambda} e^{\lambda \sigma^+}~ -~
{{2 \gamma_-} \over {\lambda^2}}\cr x^-~~ &=~~ {1 \over \lambda}
e^{\lambda \sigma^-}~ +~ {{2 \gamma_+} \over {\lambda^2}}\cr}
\eqno(4.31)$$
Thus, $\sigma^- \rightarrow - \infty$ corresponds to the lightlike
line $x^- = 2 \gamma_+/\lambda^2$ and $\sigma^+ \rightarrow - 
\infty$ to the lightlike line $x^+ = -2 \gamma_- / \lambda^2$ while
$\sigma^\pm \rightarrow \infty$ correspond to the respective
lightlike infinities.

The fluxes across both ${\cal I}_L^+$ and ${\cal I}_R^+$ are now
non-vanishing, each approaching a maximum of $\alpha \lambda^2 /2$
at early times and decreasing steadily in the far future, as $i^0$
is approached. Thus, on ${\cal I}_R^+$, for instance, we find
$$\langle T^{(\sigma)}_{++}\rangle ~ \rightarrow~ 0,~~~~~~~~
\langle T^{(\sigma)}_{+-}\rangle ~ \rightarrow~ 0 \eqno(4.32)$$
and
$$\langle T^{(\sigma)}_{--}\rangle~~ =~~ {{\alpha \lambda^2} \over
2} \left [ 1~ -~ {1 \over {\left( 1 + {{2 \gamma_+} \over \lambda}
e^{- \lambda \sigma^-} \right )^2 }}\right], \eqno(4.33)$$
and on ${\cal I}_L^+$
$$\langle T^{(\sigma)}_{--}\rangle ~ \rightarrow~ 0,~~~~~~~~
\langle T^{(\sigma)}_{+-}\rangle ~ \rightarrow~ 0 \eqno(4.34)$$
and
$$\langle T^{(\sigma)}_{++}\rangle~~ =~~ {{\alpha \lambda^2} \over
2} \left [ 1~ -~ {1 \over {\left( 1 - {{2 \gamma_-} \over \lambda}
e^{- \lambda \sigma^-} \right )^2 }}\right]. \eqno(4.35)$$
The tensors are again independent of the mass of the singularities
but depend on the soliton mass parameter. The integrated flux over
any interval is again infinite because the flux itself approaches
a steady state at early times. Of course this is a consequence of
having neglected the back reaction of the radiation on the
spacetime geometry. 

This picture is also similar to that developed in the shock wave
model. Even if quantum gravity does permit the formation of naked
singularities, they will evaporate catastrophically (``explode'')
due to the Hawking radiation. To justify this statement, one must
check the validity of the one loop picture by considering the
strength of the dilaton coupling constant at the point on the
soliton center at which the singularities are expected to detonate.
On ${\cal I}_R^+$ this is at the retarded time $x^- = 2\gamma_+/
\lambda^2$ which, when traced back corresponds to the point
$(x_\tau^+,x_\tau^-) = (1/\gamma_+ ( \Delta_0 + 2 \mu^2/ \lambda),
2\gamma_+/\lambda^2)$ on the soliton center and gives for the
dilaton coupling
$$e^\phi~~ =~~ {1 \over {\sqrt{2 \left( {{2\mu^2} \over
{\lambda^2}}~ +~ \Delta_0 \right)}}}. \eqno(4.36)$$
This is indeed small when the Bondi mass, $M_R$ is large, i.e., the
soliton mass parameter is large. On the other hand, on ${\cal
I}_L^+$ this is at the advanced time $x^+ = - 2\gamma_-/\lambda^2$
which, when traced back, corresponds to the point
$(x_\tau^+,x_\tau^-) = (- 2\gamma_-/\lambda^2, 1/\gamma_- (\Delta_0
- 2 \mu^2/ \lambda))$ on the soliton center and gives for the
dilaton coupling
$$e^\phi~~ =~~ {1 \over {\sqrt{2 \left( {{2\mu^2} \over
{\lambda^2}}~ -~ \Delta_0 \right)}}} \eqno(4.37)$$
which is again small when the Bondi mass, $M_L$, is large (or the
soliton mass parameter is large).
\vskip 0.25in

\noindent{\bf V. Formation of a Naked Singularity}

We now come to the last remaining case, the asymptotically timelike
singularity in the lower quadrant of figure V. This case is
particularly interesting in as much as it provides a bonafide model
of the ``formation'' of a naked singularity. The singularity is
formed in the future, an observer being obstructed from reaching
future null infinity, and this contrasts with our previous examples
in which the naked singularity was seen to form simultaneously with
the emergence of the soliton center. The analysis of the Hawking
radiation in this case requires somewhat different considerations
from those laid out above.

The lower branch is made up of a spacelike piece in the far future
joined smoothly to two timelike singularities on either end, being
null in the neighborhood of $\scri^-$, and intersecting $\scri_R^-$
at $x^- = -\infty, x^+ = 2 \gamma_-/\lambda^2$ and $\scri_L^-$ at
$x^+ = -\infty, x^- = - 2 \gamma_+/\lambda^2$. Classical soliton
energy entering the spacetime therefore produces a naked
singularity in the future. Accompanying the production of the
singularity is Hawking radiation, but we expect that the stress
tensor of the evaporation is regular at all points in the spacetime
except at the singularity, $\sigma = 0$. Na\"\i vely it may seem
that natural conditions are (a) there is no incoming flux of energy
other than the soliton's and (b) the Hawking radiation vanishes in
the absence of the classical soliton stress energy, i.e., when $\mu
= 0 = \Delta_0$. However, the tensor satisfying these conditions is
not regular everywhere. This is because requiring the absence of
incoming energy on $\scri^-$ (other than the soliton, condition
(a)) implies that
$$A(x^+)~~ =~~ -~ {1 \over {2 x^{+'2}}},~~~~~~~~~~ B(x^-)~~ =~~ -~
{1 \over {2 x^{-'2}}}\eqno(5.1)$$
where
$$\eqalign{x^{+'}~~ &=~~ x^+~ -~ {{2\gamma_-} \over {\lambda^2}}
\cr x^{-'}~~ &=~~ x^-~ +~ {{2 \gamma_+} \over {\lambda^2}}, \cr}
\eqno(5.2)$$
As the bracketed term on the right hand side of (4.30) is finite
everywhere, no incoming flux on $\scri^-$ implies that $\langle
T_{++} \rangle \rightarrow -  \infty$ on the lightlike line $x^{+'}
= 0$ and $\langle T_{--} \rangle \rightarrow - \infty$ on the
lightlike line $x^{-'} = 0$. This behavior is physically
unacceptable on the grounds that there is nothing special about the
spacetime along these lines and, furthermore, infinite negative
fluxes to the left and right are in violation of the positive
energy conditions. 

On the other hand the tensor that is regular everywhere within the
spacetime (except at the singularity, $\sigma = 0$) and vanishes in
the absence of the soliton is given by
$$\eqalign{\langle T_{++}\rangle~~ &=~~ T^f_{++}~ +~ T^q_{++}~~ =~~
T^f_{++}~~ -~~ \alpha \left( {{\partial^2_+ \sigma} \over \sigma}~
-~ {1 \over 2} \left[{{\partial_+ \sigma} \over \sigma}\right]^2
\right)~ -~{\alpha \over {2 x^{+2}}}\cr \langle T_{--}\rangle~~
&=~~ T^f_{--}~ +~ T^q_{--}~~ =~~ T^f_{--}~~ -~~ \alpha \left(
{{\partial^2_- \sigma} \over \sigma}~ -~ {1 \over 2}
\left[{{\partial_- \sigma} \over \sigma}\right]^2 \right)~ -~ 
{\alpha \over {2 x^{-2}}}\cr \langle T_{+-}\rangle ~~ &=~~ T^f_{+-
}~~ +~~ \alpha \partial_+ \partial_- \ln \sigma. \cr} \eqno(5.3)$$
In coordinates that are manifestly asymptotically flat
$$\eqalign{x^+~~ =~~ -~ {1 \over \lambda} e^{-\lambda \sigma^+}~~
+~~ {{2\gamma_-} \over {\lambda^2}} \cr x^-~~ =~~ -~ {1 \over
\lambda} e^{-\lambda \sigma^-}~~ -~~ {{2\gamma_+} \over
{\lambda^2}}\cr}\eqno(5.4)$$
the tensor approaches 
$$\eqalign{\langle T^{(\sigma)}_{--} \rangle~~ \rightarrow~~
0,~~~~~~~~~~ \langle T^{(\sigma)}_{+-} \rangle~~ \rightarrow~~ 0\cr
\langle T^{(\sigma)}_{++} \rangle~~ \rightarrow~~ {{\alpha
\lambda^2} \over 2} \left[ 1~~ -~~ {1 \over {\left( 1~ -~
{{2\gamma_-}\over {\lambda^2}} e^{\lambda \sigma^+} \right)^2}}
\right]\cr} \eqno(5.5)$$
on $\scri_R^-$ and 
$$\eqalign{\langle T^{(\sigma)}_{++}\rangle~~ \rightarrow~~
0,~~~~~~~~~~ \langle T^{(\sigma)}_{+-}\rangle~~ \rightarrow~~ 0\cr
\langle T^{(\sigma)}_{--}~~ \rangle \rightarrow~~ {{\alpha
\lambda^2} \over 2} \left[ 1~~ -~~ {1 \over {\left( 1~ +~
{{2\gamma_+}\over {\lambda^2}} e^{\lambda \sigma^-} \right)^2}}
\right]\cr} \eqno(5.6)$$
on $\scri_L^-$. Therefore consistency seems to require an incoming
flux of radiation across past null infinity. This flux is seen to
increase smoothly from zero on $i^-$ to the constant value $\alpha
\lambda^2 /2$ as the lightlike lines $x^- = -2\gamma_+ /\lambda^2$
and $x^+ = 2\gamma_-/\lambda^2$ are approached. The total energy
flowing into the spacetime is the integrated flux from $i^-$ to the
point $x^+ =  2\gamma_-/\lambda^2$ on the right and $x^- = - 2
\gamma_+ /\lambda^2$ on the left. This is obviously infinite and 
it is easy to see that this will occur whenever the singularity
cuts off a portion of $\scri^-$ which in turn is possible only
because of the timelike part of the singularity.

Thus the following scenario seems to emerge. Ignoring quantum
effects, the observer would enter spacetime at $i^-$ accompanied by
highly localized soliton energy and doomed to finally crash into a
singularity. Before doing so he would have to cross $x^+ =
2\gamma_-/\lambda^2$ or $x^- = -2\gamma_+ /\lambda^2$ or both. Once
he has crossed these lines he would be able to receive information
from the singularity and would realize that the cosmic censorship
hypothesis could not hold in his universe. As we have seen,
however, quantum effects will play an important role. The stress
tensor that is regular throughout the spacetime except at the
singularity itself indicates that on entering the spacetime the
observer encounters a flux of incoming energy across $\scri_L^-$
and $\scri_R^-$ and accompanying the soliton. Neglecting the back
reaction, the flux appears to increase steadily from $i^-$ leading
to an infinite total energy entering the spacetime {\it before} the
singularity becomes actually visible.
\vskip 0.25in

{\noindent \bf VI. The Back Reaction}

In as much as naked singularities are concerned, the previous case
studies fall in two categories. In the first, matter enters the
spacetime together with the formation of an asymptotically naked
singularity. The shock wave induced naked singularity and the
soliton induced singularity in the upper quadrant fall in this
category. In the second, a soliton attempts to form an
asymptotically timelike singularity in the future.  In this section
we will briefly examine the back reaction in two specific examples
of these categories: the singularity formed by a shock wave (the
first case studied) and the attempt to form a naked singularity by
an incoming soliton (the last of the cases studied earlier). 

Equation (3.1) represents the full quantum action including matter
but not including ghosts and reduces, in the weak coupling limit,
to the CGHS action (modified to include the anomaly term). After
the field redefinitions in (3.6), the action may be written in
terms of the fields $\chi$ and $\Omega$ as follows,
$$\eqalign{S~~ =~~ \int d^2x & \left[ - \partial_+ \chi \partial_-
\chi~ +~ \partial_+\Omega \partial_- \Omega~ +~ \sum_i \partial_+
f_i \partial_- f_i \right. \cr &~~~~ \left. +~ {1 \over 2} \left(
\Lambda~ +~ 4 \mu^2 U({\vec f}) \right) e^{{\sqrt{2 \over
{|\alpha|}}} (\Omega \mp \chi)} \right]\cr}\eqno(6.1)$$
plus a ghost action. From the above action follow the equations of
motion for $\chi$ and $\Omega$ as well as the constraints
$$\eqalign{\partial_+ \partial_- \Omega~~ &=~~ \pm~ \partial_+
\partial_- \chi~~ =~~ {\sqrt{2 \over {|\alpha|}}} \left({\Lambda
\over 4}~ +~ \mu^2 U({\vec f})\right) e^{{\sqrt{2 \over |\alpha|}}
(\Omega \mp \chi))}\cr \partial_+ \partial_- f_i~~ &=~~ \mu^2
e^{{\sqrt{2 \over \alpha}} (\Omega \mp \chi)} {{\partial U({\vec
f})} \over {\partial f_i}}\cr {\cal T}_{\pm\pm}~~ &=~~ \mp~ {1
\over 2} \partial_\pm \chi \partial_\pm \chi~ \pm~ {1 \over 2}
\partial_\pm \Omega \partial_\pm \Omega~ +~ {1 \over 2} \sum_i
\partial_\pm f_i \partial_\pm f_i~ +~ {\sqrt{{|\alpha|} \over 2}}
\partial_\pm^2 \Omega~ +~ t_{\pm\pm} \cr}\eqno(6.2)$$
where $t_{\pm\pm}$ is the ghost stress energy and $U({\vec f})$ is
the matter field potential, being zero for the shock wave, and 
$$U({\vec f})~~  =~~ \left( \prod_{i=1}^{N/2} \cos f_i
\prod_{j=N/2+1}^N \cosh f_j~ -~ 1\right) \eqno(6.3)$$
for the soliton solutions described in sections IV and V.

We are now in a position to derive the exact solutions to quantum
dilaton gravity in two dimensions. Evidently we may choose
$\chi(x^+,x^-) = \pm \Omega(x^+,x^-)$ and the solution which obeys
the constraints ${\cal T}_{\mu\nu} = 0$, with $t_{\pm\pm}=0$, may
be written down directly in comparison with the solutions given in
the previous sections for the weak coupling limit. In all cases
$$\Omega(x^+,x^-)~~ =~~ \pm~ \chi(x^+,x^-)~~ =~~ -~ {\sqrt{2 \over
{|\alpha|}}} \Gamma(x^+,x^-)\eqno(6.4)$$
where
$$\Gamma(x^+,x^-)~~ =~~ \left\{\matrix{-~ {\Lambda \over 4} x^+ x^-
~ -~ a (x^+ - x^+_0) \Theta(x^+ - x^+_0) \cr\cr -~{\Lambda \over 4}
x^+ x^-~ -~ 2 \ln\cosh(\Delta - \Delta_0)}\right. \eqno(6.5)$$
where the first is for the incoming shock wave and the second for
the sine-Gordon solitons.

The solutions are the same as in the CGHS theory except that now
they are given in terms of $\chi$ and $\Omega$ which incorporate
all the quantum effects. In the limit of small coupling, equation
(3.9) implies that $\chi = \Omega \sim - e^{-2\rho} \sim - e^{-
2\phi}$. This is to be expected of course and it gives the
classical solution admitting the singularities that have already
been described, but the singularities appear in the strong coupling
regime and the strong coupling expansion of (3.6) and (3.7) should
have been used. Moreover, the quantum matter stress energy contains
a boundary condition dependent piece (the contribution from the
conformal anomaly has already been accounted for) so that, if the
constraint ${\cal T}_{\mu\nu} = 0$ is to be maintained, the ghost
contribution must cancel this contribution and therefore cannot be
zero. Therefore the solutions in (6.4) and (6.5) need to be
modified to account for the boundary condition dependence of the
Hawking radiation. For definiteness, we will take $\alpha > 0$ ($N
< 24$) in what follows.
Referring back to equation (3.6) we are reminded of the fact that
the fields, $\chi$ and $\Omega$, are defined in terms of the
dilaton and conformal factor but that this definition depends
rather strongly on the way in which the field space metric,
$G_{ab}$, and the tachyon, $T(X)$, are renormalized. There are many
possibilities and we shall consider but one such here, taking 
$$h_1(\phi)~~ =~~ -~ {{3\alpha} \over 4} e^{2\phi},~~~~~~~~~~~~
h_2(\phi)~~ =~~ -\alpha e^{2\phi} \eqno(6.6)$$
which amounts to a renormalization of the field space metric but
not the tachyon and has the advantage of making the square-rooted
term in the definition of $P(\phi)$ a perfect square. This model is
similar to that of Russo, Susskind and Thorlacius,${}^8$ and it
gives
$$\eqalign{P(\phi)~~ &=~~ 2 {\sqrt{2 \over \alpha}} e^{-2\phi}
\left[ 1~ -~ {\alpha \over 2} e^{2\phi} \right] \cr \chi~~  &=~~ 
\int d\phi P(\phi)~~ =~~ -~ {\sqrt{2 \alpha}} \left[ \phi~ +~
{{e^{-2\phi}} \over \alpha} \right]\cr \Omega~~ &=~~
{\sqrt{2\alpha}} \left[\rho~ -~ 2 \phi~ -~ {{e^{-2\phi}} \over
\alpha} \right]\cr} \eqno(6.7)$$
so that the solution $\chi = \Omega$ implies that $\rho = \phi$ and
$\Omega-\chi = {\sqrt{2\alpha}}(\rho-\phi)$.

Now there is a certain arbitrariness in the choice of the ghost
stress energy, which is associated with the freedom to choose the
boundary conditions. One possible way to fix this is to prescribe
its form in such a way as to be consistent with the semi-classical
Hawking radiation on $\scri^+$.  This was seen to approach a
constant at early retarded times and decay to zero in the far
future, so that the integrated flux, the total energy leaving the
spacetime from any point $x^-$ on $\scri^+_R$ to future timelike
infinity, $i^+$, is then
$$\int_{\sigma^-}^\infty d\sigma^- \langle T^{(\sigma)}_{--}
\rangle~~ =~~ {{\alpha \lambda} \over 2} \left[ \ln \left(1 + {a
\over \lambda} e^{-\lambda \sigma^-} \right)~ +~ {1 \over {\left(1
+ {\lambda \over a} e^{\lambda \sigma^-}\right)}} \right]
\eqno(6.8)$$
in the double null coordinates of (2.20).

Requiring that the solution reproduce the Hawking boundary
conditions means that the Bondi mass, measured on $\scri^+$, should
{\it decrease} along $\scri^+_R$ at the same rate as the flux of
outgoing Hawking radiation, either becoming infinitely negative by
the time $i^+$ is approached if the initial mass is finite, or
possessing an infinite initial mass if the final mass (on $i^+$) is
finite. Moreover, to be consistent with the semi-classical
calculation, the mass on $\scri^+_L$ should be constant, and the
total energy along both null infinities must be conserved. The
following choice for the ghost stress energy 
$$t_{++}~~ =~~ {\alpha \over {2x^{+2}}},~~~~~~~~~~ t_{--}~~ =~~
{\alpha \over {2x^{-'2}}}. \eqno(6.9)$$
where 
$$x^{-'}~~ =~~ x^-~ -~ {a \over \lambda^2}\Theta(x^+-x^+_0)
\eqno(6.10)$$
satisfies all these conditions. It leads to the solution for
$\chi(x^+,x^-)$ obeying the modified constraints
$$\Omega~~ =~~ \chi~~ =~~ -~ {\sqrt{2 \over \alpha}} \left(
\lambda^2 x^+ x^-~ -~ a(x^+~ -~ x^+_0)\Theta(x^+-x^+_0)~ -~ {\alpha
\over 2} \ln (\lambda^2 x^+ x^{-'})\right).\eqno (6.11)$$
In order to compute the Bondi mass of the solution following the
arguments of section II,  we need a reference solution,
$\Omega^{(0)}$, and an expansion about it of the form $\Omega =
\Omega^{(0)} + \delta \Omega$.  It is natural to take this
reference to be the linear dilaton solution, which is the value of
$\Omega$ in the absence of the incoming shock wave,
$$\Omega^{(0)}~~ =~~ -~ {\sqrt{2 \over \alpha}} \left[ \lambda^2
x^+ x^-~ -~ {\alpha \over 2} \ln (\lambda^2 x^+ x^-) \right]
\eqno(6.12)$$
Comparing it with the full solution in (6.11) we then get 
$$\delta \Omega~~ =~~ -~ {\sqrt{2 \over \alpha}} \left[ {M \over
\lambda}~ -~ ax^+~ -~ {\alpha \over 2} \ln ({{x^{-'}} \over {x^-}})
\right] \eqno(6.13)$$
whereupon, using the stress tensor in (6.2), the charge evolving
the system along $\scri^+_R$ are seen to be given by 
$$\eqalign{Q^-~~ &=~~ \lambda {\sqrt{\alpha \over 2}} \left[-~
\delta \Omega~ +~ x^+ \partial_+ \delta \Omega~ +~ x^{-'} \partial
_- \delta \Omega \right]_{\scri_R^+}\cr&=~~ \lambda \left[ {M \over
\lambda}~ +~ {\alpha \over 2} \ln\left({{x^-} \over {x^{-'}}}
\right)~ +~ {\alpha \over 2} \left( 1 - {{x^{-'}} \over {x^-}}
\right) \right] \cr}\eqno(6.14)$$
First we remark that the constant $M$ is the Bondi mass measured by
an observer on $i^+$. The charge is a constant on $\scri^+_L$, and
infinite in the retarded past on $\scri^+_R$, at $x^-=a/\lambda^2$.
This behavior is required to reproduce the semi-classical Hawking
radiation given in section II: an evaporating singularity that
gives up an infinite amount of energy to end up with a finite mass
must have been infinitely massive to begin with. The total energy,
including the integrated flux, is conserved. This is easy to see by
rewriting the expression in (6.14) in terms of the coordinates
$\sigma^\pm$,
$$Q^-~~ =~~ M(\sigma^-)~~ =~~ M~ +~ {{\alpha \lambda} \over 2}
\left[ \ln \left(1 + {a \over \lambda} e^{-\lambda \sigma^-}
\right)~ +~ {1 \over {\left(1 + {\lambda \over a} e^{\lambda
\sigma^-}\right)}} \right]. \eqno(6.15)$$
and one recognizes the last two terms on the right hand side as the
integrated flux over the interval $\sigma^-$ to $\infty$ on
$\scri^+_R$, given in (6.8). In fact, the Hawking flux across
$\scri^+_R$ may be calculated as the negative of the rate of change
of the mass in (6.15),
$$-~ {{dM(\sigma^-)} \over {d\sigma^-}}~~ =~~ {{\alpha\lambda^2}
\over 2}~ \left[1~ -~ {1 \over {(1+{a \over \lambda} e^{- \lambda
\sigma^-})^2}}\right],\eqno(6.16)$$
which is the proper way to think of the Hawking effect in the full
theory.

When $M=0$, (6.11) represents a shifted dilaton vacuum. When $M
\neq 0$, a singularity is formed at $\chi = \chi_{\rm min}$,
$${\alpha \over 2} \ln \left({{2e} \over \alpha}\right)~~ =~~
\lambda^2 x^+ x^{-'}~ +~ {M \over \lambda}~ -~ {\alpha \over 2} \ln
( \lambda^2 x^+ x^{-'}). \eqno(6.17)$$
Clearly there are no solutions for $M > 0$ because the minimum of
the sum of the first and last term in (6.17) is precisely equal to
the left hand side. For $M<0$ there are two real solutions. These
are given by
$$\lambda^2 x^+ x^{-'}~~ =~~ -~ {\alpha \over 2} {\cal W}(k,-
e^{{{2M} \over {\alpha\lambda}} -1}).\eqno(6.18)$$
where $W(k,x)$ is Lambert's function${}^{17}$ and $k=0,-1$. The
branch corresponding to $k=0$ is the principal branch and is the
one that is analytic at $x=0$ ($M \rightarrow -\infty$).

As $M$ is the Bondi mass measured at future timelike infinity and
on $\scri^+_L$, the only singularities possible within the full
quantum theory are {\it spacelike} and would have negative Bondi
masses on $\scri^+_L$. Measured on $\scri^+_R$, however, they would
begin with a positive mass in the past and rapidly give up energy
over retarded time to turn eventually into spacelike negative
energy singularities. The Penrose diagram is displayed in figure
VII (where the physical regions in which $\sigma > 0$ are denoted
by I and III). For positive $M$, quantum effects have prevented the
singularity from forming. For negative $M$, quantum effects turn
the naked singularity into a black hole in region III and a white
hole in region I. (Even classically the singularity is a black hole
for negative $M$, but of course negative $M$ is physically not
allowed.)

The next case study we examine is that of section V, where the
incoming soliton attempts to form an asymptotically naked
singularity in the future. Two possible choices of the ghost
contribution to the full stress energy were discussed. The first
choice involved no incoming Hawking radiation and led to an
negative infinite flux of energy on the future horizon in the semi-
classical approximation. The second choice avoided this problem by
allowing an influx of Hawking energy across past null infinity. We
will now discuss the choices of ghost stress energy that correspond
to each of these boundary conditions.

If there is no incoming Hawking flux at past null infinity, the
appropriate choice for the ghost stress energy is
$$t_{\pm\pm}~~ =~~ {\alpha\over {2x^{\pm 2}}} \eqno(6.19)$$
which leads to the following solution
$$\Omega~~ =~~ \chi~~ =~~ -~ {\sqrt{2\over \alpha}}\left(\lambda^2
x^+ x^-~ -~ 2\ln\cosh(\Delta-\Delta_0)~ -~ {\alpha \over 2} \ln
(\lambda^2 x^+ x^-) \right). \eqno(6.20)$$
To see that this is indeed correct, consider the Bondi mass as
measured by an observer on $\scri^-$. For example, choosing 
$\Omega^{(0)}$ (the vacuum) as before, in (6.12), on $\scri^-_R$
($x^- \rightarrow - \infty$) we have
$$\eqalign{\delta \Omega~~ &=~~ -~ {\sqrt{2 \over \alpha}} \left[ -
~ 2 \ln \cosh (\Delta - \Delta_0) \right] \cr &\sim~~ 2 {\sqrt{2
\over \alpha}} \left[ \gamma_+ x^+~ +~ \gamma_- x^-~ -~ \Delta_0~ -
~ \ln 2 \right]\cr} \eqno(6.21)$$
so that the charge evolving the system along $\scri^-_R$ is
constant:
$$\eqalign{Q^+~~ &=~~ \lambda {\sqrt{2 \over \alpha}} \left[ -
\delta \Omega~ +~ x^{+'} \partial_+ \delta \Omega~ +~ x^-
\partial_- \delta \Omega \right] |_{\scri^-_R}\cr &=~~ 2\lambda
\left[ {{2\mu^2} \over {\lambda^2}}~ +~ \Delta_0~ +~ \ln 2
\right]~~ =~~M_R \cr}\eqno(6.22)$$
where
$$x^{+'}~~ =~~ x^+~ -~ {{2\gamma_-} \over {\lambda^2}}.$$
Likewise, on $\scri^-_L$, the Bondi mass is calculated to be
$$\eqalign{Q^-~~ &=~~ \lambda {\sqrt{2 \over \alpha}} \left[ -
\delta \Omega~ +~ x^+ \partial_+ \delta \Omega~ +~ x^{-'}
\partial_- \delta \Omega \right] |_{\scri^-_L}\cr &=~~ 2\lambda
\left[ {{2\mu^2} \over {\lambda^2}}~ -~ \Delta_0~ +~ \ln 2
\right]~~ =~~ M_L\cr} \eqno(6.23)$$
where
$$x^{-'}~~ =~~ x^-~ +~ {{2\gamma_+} \over {\lambda^2}}$$
in the notation of sections IV and V.

The singularity occurs at
$${\alpha\over 2} \ln \left({{2e} \over \alpha}\right)~~ =~~
\lambda^2 x^+ x^-~ -~ 2\ln\cosh(\Delta-\Delta_0)~ -~ {\alpha \over
2} \ln (\lambda^2 x^+ x^-). \eqno(6.24)$$
This curve is obtained quite easily in parametrized form (using
$\Delta$ as a parameter) and can be written in terms of the
Lambert's ${\cal W}$ function as follows
$$\eqalign{x^-~~ &=~~ {1 \over {2 \gamma_-}}\left[\Delta~ \mp~
{\sqrt{\Delta^2~ -~ 2\alpha {{\mu^2}\over{\lambda^2}} {\cal
W}[k,f(\Delta)]}} \right]\cr x^+~~ &=~~ {{\Delta - \gamma_- x^-}
\over {\gamma_+}}\cr f(\Delta)~~ &=~~ -~ \exp\left[-~ 1~ -~ {4
\over \alpha} \ln \cosh (\Delta - \Delta_0) \right]\cr}
\eqno(6.25)$$
for $k=0,-1$. The range of $\Delta$ is $(-\infty, \infty)$ and the
singularities are displayed in figure VIII. An observer entering
spacetime in region III will see a naked singularity before
inevitably crashing into it. Region II is unphysical. An observer
in region I would emerge from a white hole.

If, on the other hand, we do allow for incoming Hawking radiation
on $\scri^-$, the appropriate choice for the ghost stress energy is
$$t_{\pm\pm}~~ =~~ {\alpha\over {2x^{\pm' 2}}} \eqno(6.26)$$
which gives the following solution for $\Omega$
$$\Omega~~ =~~ -~ {\sqrt{2\over \alpha}} \left[ \lambda^2 x^+ x^-~
-~ 2 \ln \cosh (\Delta - \Delta_0)~ -~ {\alpha \over 2} \ln
\lambda^2 x^{+'} x^{-'} \right]. \eqno(6.27)$$
That this solution indeed represents incoming Hawking radiation on
$\scri^-$ is checked by calculating the Bondi mass of the spacetime
there. One gets
$$\delta \Omega~~ =~~ -~ {\sqrt{2 \over \alpha}} \left[ -~ 2 \ln
\cosh (\Delta - \Delta_0)~ -~ {\alpha \over 2} \left( \ln {{x^{+'}}
\over {x^+}}~ +~ \ln {{x^{-'}} \over {x^-}} \right) \right].
\eqno(6.28)$$
Thus, on $\scri_R^-$ ($x^- \rightarrow -\infty$), 
$$\delta \Omega~~ \sim~~ 2 {\sqrt{2 \over \alpha}} \left[\gamma_+
x^+~ +~ \gamma_- x^-~ -~ \Delta_0~ -~ \ln 2~ +~ {\alpha \over 4}
\ln {{x^{+'}} \over {x^+}} \right] \eqno(6.29)$$
giving,
$$Q^+~~ =~~ \lambda \left[{M_R\over \lambda}~ +~ {\alpha \over 2}
\ln {{x^+} \over {x^{+'}}}~ +~ {\alpha \over 2} \left( 1~ -~
{{x^{+'}} \over {x^+}} \right) \right] \eqno(6.30)$$
and, similarly, on $\scri_L^-$, 
$$Q^-~~ =~~ 2 \lambda \left[{M_L\over \lambda}~ +~ {\alpha \over 2}
\ln {{x^-} \over {x^{-'}}}~ +~ {\alpha \over 2} \left( 1~ -~ {{x^{-
'}} \over {x^-}} \right) \right] \eqno(6.31)$$
These are, of course, the Bondi masses measured on the respective
past null infinities. Changing to asymptotically flat coordinates
defined by
$$\eqalign{x^+~ &=~ -~ {1 \over \lambda} e^{-\lambda \sigma^+}~ +~
{{2\gamma_-} \over {\lambda^2}}\cr x^-~ &=~ -~ {1 \over \lambda}
e^{-\lambda \sigma^-}~ -~ {{2\gamma_+} \over {\lambda^2}}\cr}
\eqno(6.32)$$
one finds that
$$\eqalign{M_R(\sigma^+)~~ &=~~ M_R~ +~ {{\alpha \lambda} \over 2}
\left[\ln \left(1~ -~ {{2\gamma_-} \over \lambda} e^{\lambda
\sigma^+} \right)~ +~ \left({1 \over {1~ -~ {\lambda \over
{2\gamma_-}} e^{-\lambda \sigma^+}}}\right) \right]\cr M_L(\sigma^-
)~~ &=~~ M_L~ +~ {{\alpha \lambda} \over 2} \left[ \ln \left(1~ +~
{{2\gamma_+} \over \lambda} e^{\lambda \sigma^-} \right)~ +~
\left({1 \over {1~ +~ {\lambda \over {2\gamma_+}} e^{-\lambda
\sigma^-}}}\right)\right]\cr} \eqno(6.33)$$
Therefore,
$$\eqalign{{{d M_R(\sigma^+)} \over {d\sigma^+}}~~ &=~~ \langle
T^{(\sigma)}_{++} \rangle\cr {{d M_L(\sigma^-)} \over {d\sigma^-
}}~~ &=~~ \langle T^{(\sigma)}_{--} \rangle\cr} \eqno(6.34)$$
on past null infinity, where $\langle T^{(\sigma)}_{\pm\pm}
\rangle$ were given in (5.5) and (5.6).

We have not been able to obtain a closed form expression for the
singularity curve in this case. Yet, one can analyze this curve in
the asymptotic regions where, for example on $\scri^-_R$, its
equation reads
$${\alpha\over 2} \ln \left({{2e} \over \alpha}\right)~~ =~~
\lambda^2 x^{+'} {\tilde x}^-~ +~ {M \over \lambda}~ -~ {\alpha
\over 2} \ln (\lambda^2 x^{+'} x^{-'}). \eqno(6.35)$$
with
$${\tilde x}^-~~ =~~ x^-~ -~ {{2\gamma_+} \over {\lambda^2}}.$$
As $x^- \rightarrow -~ \infty$, this equation is similar to the
corresponding one for the shock wave (equation 6.17) and the
conclusions are the same. There is no solution for $M>0$. $M=0$ is
the (shifted) vacuum and, when $M<0$, the singularity is
(asymptotically) spacelike, intersecting $\scri^-_R$ at $x^+ = 0$.
Obviously, a parallel conclusion can be drawn for the singularity
on $\scri^-_L$ and a negative mass singularity would intersect
$\scri^-_L$ at $x^- = 0$. It looks as if quantum effects prevent
the singularity from forming if $M>0$ and change it to a black hole
if $M<0$.
\vskip 0.25in

{\noindent \bf VII. Discussion}

This paper represents an attempt to explore the consequences of a
violation of the classical cosmic censorship conjecture. One may
imagine that Einstein's theory predicts that some reasonable
physical systems will collapse to form naked singularities. It is
generally, but not universally, believed that the breakdown due to
the formation of a timelike (naked) singularity is more serious
than the breakdown due to the formation of a spacelike (black hole)
singularity. In either case one does expect quantum effects to play
an important role. Since string theory is the only available theory
that gives a consistent quantum theory of gravity, it represents
the scenario in which problems involving gravitational collapse to
a naked singularity should properly be explored. Our work
represents an attempt to arrive at an understanding of the details
of gravitational collapse on the classical and the quantum levels. 

On the classical level, and in the case of the soliton induced
singularities, the changing character of each singularity is
fascinating.  In the same spacetime the singularity may alternate
and smoothly change its nature from being spacelike to timelike and
back to spacelike. Quite possibly this type of behavior may be
typical of the collapse of a chaotic distribution of matter. The
induced radiation accompanying the collapse is in all cases very
significant and implies that a major revision of the classical
conclusions is necessary.

On the quantum level, we have analyzed the problem first semi-
classically as this is the basis of our physical intuition.  Here
already we found indications of a weak cosmic censorship
hypothesis. If a naked singularity forms, the total energy radiated
will exceed the mass of the incoming physical field. In the case of
a black hole the energy radiated will build up slowly but in the
case of a naked singularity the radiated energy builds up rapidly.
Thus, even at this stage one might conjecture that when the
collapse of a physical system is predicted by Einstein's theory to
lead to a naked singularity, quantum effects would induce an
explosive burst of radiation such that the back reaction would step
in to prevent the singularity from actually forming. The quantum
treatment including the back reaction seems to verify this
conjecture, if attention is restricted to positive (Bondi) mass
singularities and boundary conditions that obey the weak energy
conditions in the semi-classical approximation.

The tantalizing question is then: do phenomena such as these exist
in nature and do they lead to observable astrophysical effects?
\vskip 0.25in

\noindent{\bf Acknowledgements}

This work was supported in part by  NATO under contract number CRG
920096. C. V. acknowledges the partial support of the {\it Junta
Nacional de Investiga\c{c}\~ao Cient\'\i fica e Tecnol\'ogica}
(JNICT) Portugal under contracts number CERN/S/FAE/1037/95 and
CERN/S/FAE/1030/95, and L.W. acknowledges the partial support of
the U. S. Department of Energy under contract number DOE-FG02-
84ER40153.
\vfill
\eject

\noindent {\bf Figure Captions:}
\vskip 0.2in

{\item{}}{\bf Figure I:} Penrose diagram of the naked singularity
represented by (2.6)

{\item{}}{\bf Figure II:} Penrose diagram of the naked singularity
which appears simultaneously with an incoming $f-$ shock wave
(equation 2.15).

{\item{}}{\bf Figure III:} The Kruskal diagram for $\Lambda = 4
\lambda^2 > 0$ and $\Delta_0 > 0$. Regions I \& III are physical
($\sigma > 0)$. The singularities are spacelike at null infinity
and joined smoothly along the soliton center by a timelike
singularity. The figure was drawn for the following parameter
values: $\mu^2 = 1$, $\lambda^2 = 4$, $\gamma_+ = \sqrt{3}$,
$\gamma_- = -1/\sqrt{3}$, $\Delta_0 = 0.1$ and $\alpha = 1/24\pi$. 

{\item{}}{\bf Figure IV:} The Kruskal diagram for $\Lambda =
4\lambda^2 > 0$ and $\Delta_0 = 0$. Regions I \& III are physical
($\sigma > 0$). Both singularities are spacelike so that a black
hole and a white hole are seen to intersect at the origin. The
figure was drawn for the following parameter values: $\mu^2 = 1$,
$\lambda^2 = 4$, $\gamma_+ = \sqrt{3}$, $\gamma_- = -1/\sqrt{3}$
and $\alpha = 1/24\pi$. 

{\item{}}{\bf Figure V:} The Kruskal diagram for $\Lambda = -
4\lambda^2 < 0$ and $\Delta_0 > 0$. Regions II \& IV are physical
($\sigma > 0)$. The singularities are timelike at null infinity and
joined smoothly at the soliton center. The figure was drawn for the
following parameter values: $\mu^2 = 1$, $\lambda^2 = 0.25$,
$\gamma_+ = \sqrt{3}$, $\gamma_- = -1/\sqrt{3}$, $\Delta_0 = 5.0$
and $\alpha = 1/24\pi$. 

{\item{}}{\bf Figure VI:} The Kruskal diagram for $\Lambda = -4
\lambda^2 < 0$ and $\Delta_0 = 0$. Regions II \& IV are physical
($\sigma > 0)$. Two timelike singularities intersect at the origin.
The figure was drawn for the following parameter values: $\mu^2 =
1$, $\lambda^2 = 0.25$, $\gamma_+ = \sqrt{3}$, $\gamma_- = -
1/\sqrt{3}$ and $\alpha = 1/24\pi$.

{\item{}}{\bf Figure VII:} The Penrose diagram for the negative
mass (on $\scri^+_L$) singularity formed in the full quantum theory
by an incoming $f-$shock wave. Regions I and III are physical.

{\item{}}{\bf Figure VIII:} The Kruskal diagram for the $k=0$
branch of the singularity formed by an incoming soliton in the full
quantum theory. The boundary conditions are no incoming radiation
on $\scri^-$. The figure was drawn for the following parameter
values: $\Delta_0 = 1$, $\lambda^2=1$, $\gamma_+ = \sqrt{3}$,
$\gamma_- = -1/\sqrt{3}$ and $\alpha=23/6$. Regions I and III are
physical.

\vfill\eject

\noindent{\bf References}

{\item{1.}}S. W. Hawking , Comm. Math. Phys. {\bf 43} (1975) 199;
J. B. Hartle and S. W. Hawking, Phys. Rev. {\bf D13} (1976) 2188;
B.S. DeWitt, Phys. Rep. 19 (1975) 295; W. Israel, Phys. Lett. {\bf
A 57} (1976) 107; W. G. Unruh, Phys. Rev. {\bf D14} (1976) 840;

{\item{2.}}S. M. Christensen and S. A. Fulling, Phys. Rev. {\bf
D15} (1977) 2088. 

{\item{3.}}E. Witten Phys. Rev. {\bf D44} (1991) 314; G. Mandal, M.
M. Sengupta, and S. R. Wadia, Mod. Phys. Letts. {\bf A6} (1991)
1685.

{\item{4.}}A. Sen, Nucl. Phys. {\bf B434} (1995) 421; {\it ibid},
Nucl. Phys. {\bf B447} (1995) 62.

{\item{5.}}G. Gibbons and K. Maeda, Nucl. Phys. {\bf B298} (1988)
741; D. Garfinkle, G. Horowitz and A. Strominger, Phys. Rev. {\bf
D43} (1991) 3140; G. Horowitz, in {\it String Theory and Quantum
Gravity, 1991}, Proceedings of the Trieste Summer School 1991, eds.
J. Harvey, R. Iengo, K. Narain, S. Randjbar-Daemi and H. Verlinde,
World Scientific (1991).

{\item{6.}}C. Callan, S. B. Giddings, J. Harvey and A. Strominger,
Phys. Rev. {\bf D45} (1992) 1005; 

{\item{7.}}S. de Alwis, Phys. Letts. {\bf B289} (1992) 282; Phys.
Rev. {\bf D46} (1992) 5429; A. Bilal and C. Callan, Nucl. Phys.
{\bf B394} (1993) 73; 

{\item{8.}}J. G. Russo, L. Susskind and L. Thorlacius, Phys. Rev.
{\bf D46} (1993) 3444; Phys. Rev. {\bf D47} (1993) 533; T. M.
Fiola, J. Preskill, A. Strominger and S. Trivedi, Phys. Rev. {\bf
D50} (1994) 3987; A. Strominger and L. Thorlacius, Phys. Rev. {\bf
D50} (1994) 5177; J. Polchinski and A. Strominger, Phys. Rev. {\bf
D50} (1994) 7403; S. P. de Alwis and D. A. MacIntire, Phys. Letts.
{\bf B344} (1995) 110; D. Louis Martinez and G. Kunstatter, Phys.
Rev. {\bf D52} (1995) 3494.

{\item{9.}}Sukanta Bose, L. Parker and Yoav Peleg, Phys. Rev. {\bf
D52} (1995) 3512; {\it ibid} Phys. Rev {\bf D53} (1996) 5708; {\it
ibid}, Phys. Rev. {\bf D53} (1996) 7089; {\it ibid}, Phys. Rev.
Letts. {\bf 76} (1996) 861; A. Fabri and J. G. Russo, Phys. Rev.
{\bf D53} (1996) 6995.

{\item{10.}}P. S. Joshi, {\it Global Aspects in Gravitation and
Cosmology}, Clarendon Press, Oxford, 1993. Chapter 6.
{\item{11}}Sundeep K. Chakrabarti and Pankaj S. Joshi, Int. J. Mod.
Phys. {\it D3} (1994) 647; P.S. Joshi and T. P. Singh, Phys. Rev.
{\bf D51} (1995) 6778; I. H. Dwivedi and P.S. Joshi, Comm. Math.
Phys. {\bf 166} (1994) 117; T. P. Singh and P. S. Joshi, Class.
Quant. Grav. {\bf 13} (1996) 559. F. I. Cooperstock, S. Jhingan,
P.S. Joshi and T. P. Singh, {\it Negative Pressures and Naked
Singularities in Spherical Gravitational Collapse}, gr-qc/9609051.

{\item{12.}}R. Penrose, Riv. Nuovo Cimento {\bf 1} (1969) 252; in
{\it General Relativity, An Einstein Centenary Survey}, ed. S. W.
Hawking and W. Israel, Cambridge Univ. Press, Cambridge, England,
(1979) 581. In its original form, the Cosmic Censorship Hypothesis
(CCH) essentially states that: {\it no physically realistic
collapse, evolving from a well posed initial data set and
satisfying the dominant energy condition, results in a singularity
in the causal past of null infinity}. There is also a strong
version of the CCH which states that: {\it no physically realistic
collapse leads to a locally timelike singularity}.

{\item{13.}}Cenalo Vaz and Louis Witten, Phys. Letts. {\bf B327}
(1994) 29; Class. Quant. Grav. {\bf 12} (1995) 2607; Class. Quant.
Grav. {\bf 13} (1996) L59.

{\item{14.}}Hak-Soo Shin and Kwang-Sup Soh, Phys. Rev. {\bf D52}
(1995) 981 

{\item{15.}}S. Weinberg, {\it Gravitation and Cosmology}, John
Wiley and Sons, Inc. (1972).

{\item{16.}}For a review, see A. Strominger in Les Houches Lectures
on Black Holes: Lectures Presented at the 1994 Summer School, {\it
Fluctuating Geometries in Statistical Mechanics and Field Theory},
hep-th/9501071.

{\item{17.}}R. M. Corless, G. H. Gonnet, D.E.G. Hare and D. J.
Jeffrey, ``On Lambert's ${\cal W}$ function'', Technical Report Cs-
93-03, Dept. of Computer Science, Univ. of Waterloo, Canada
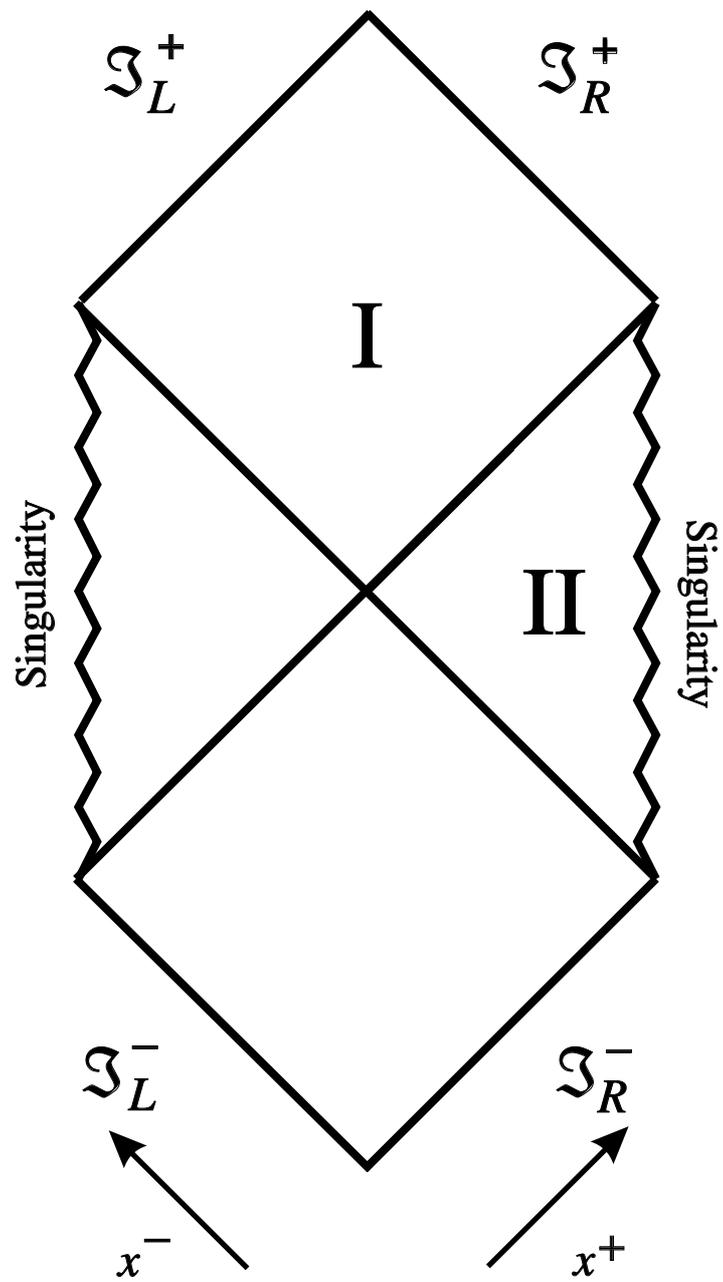
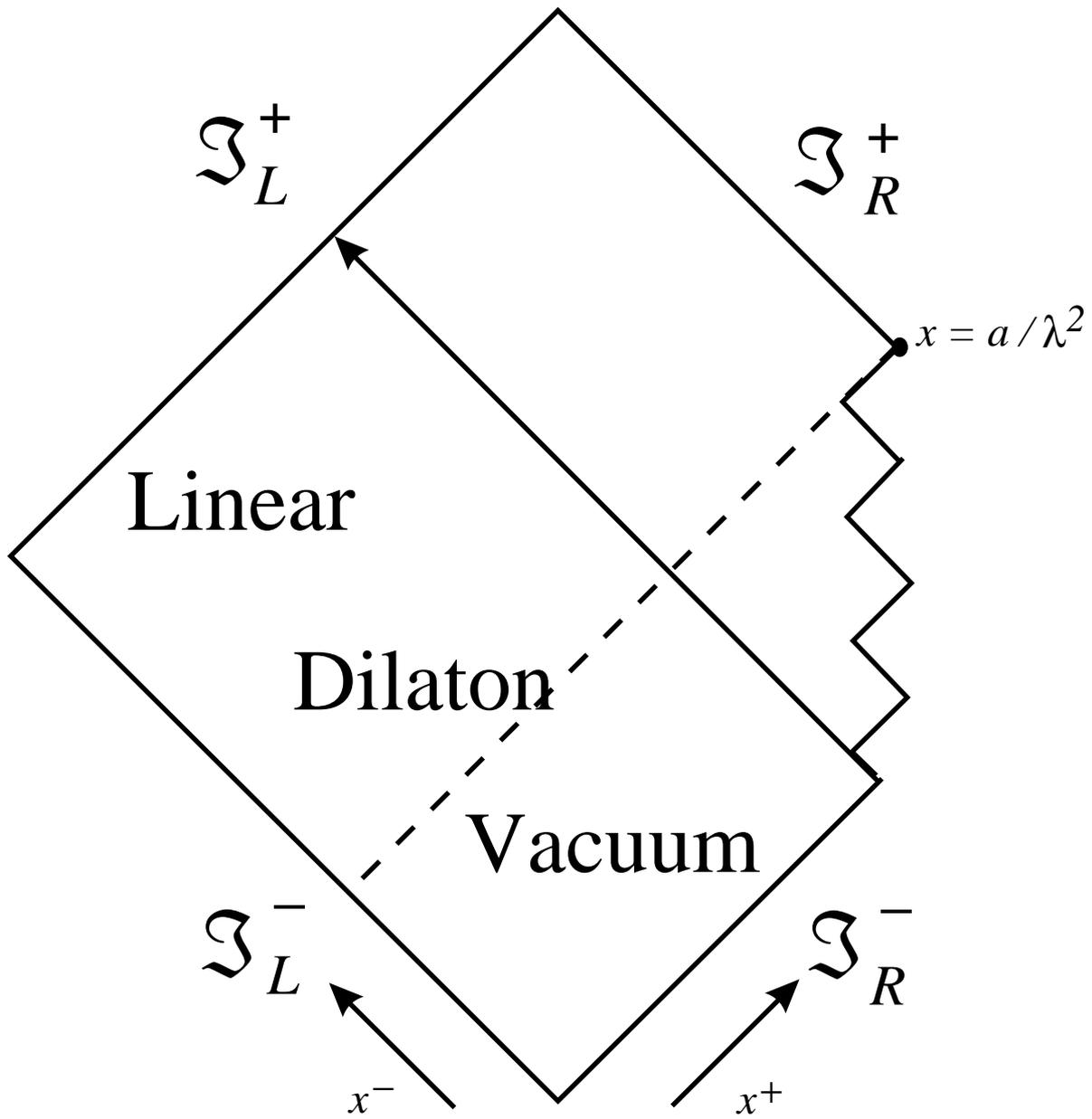
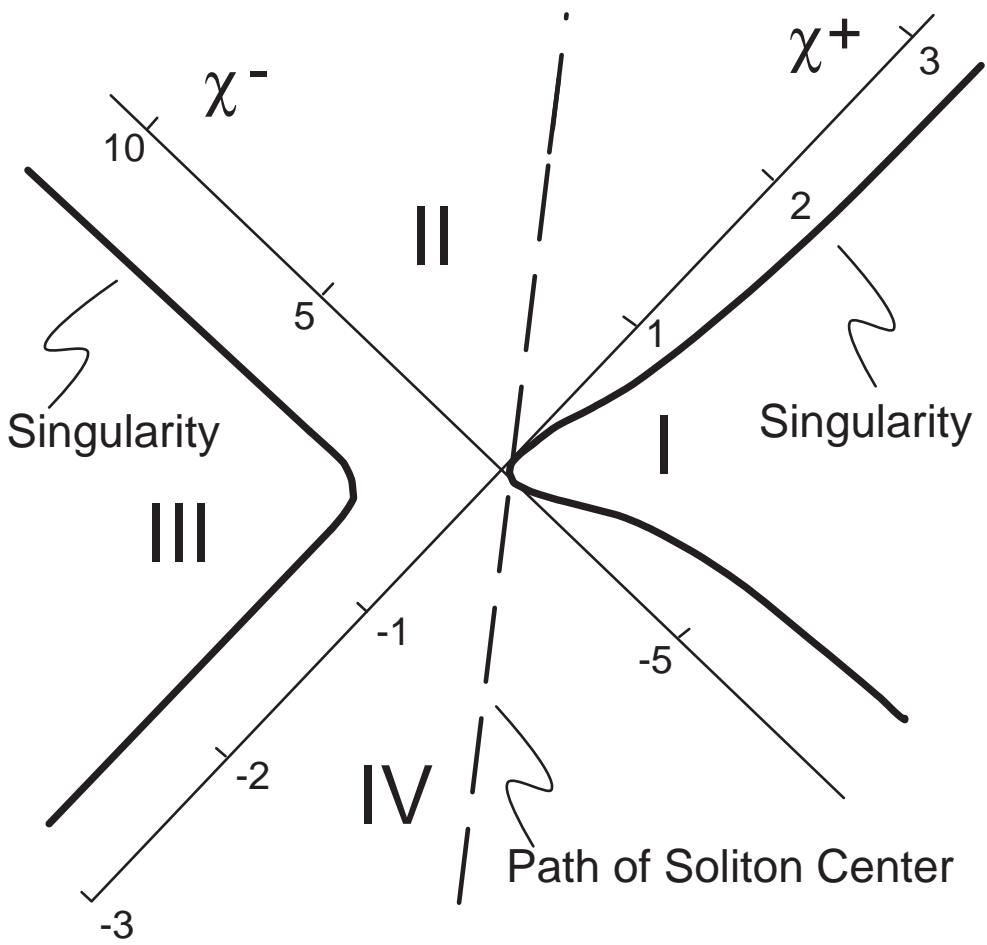
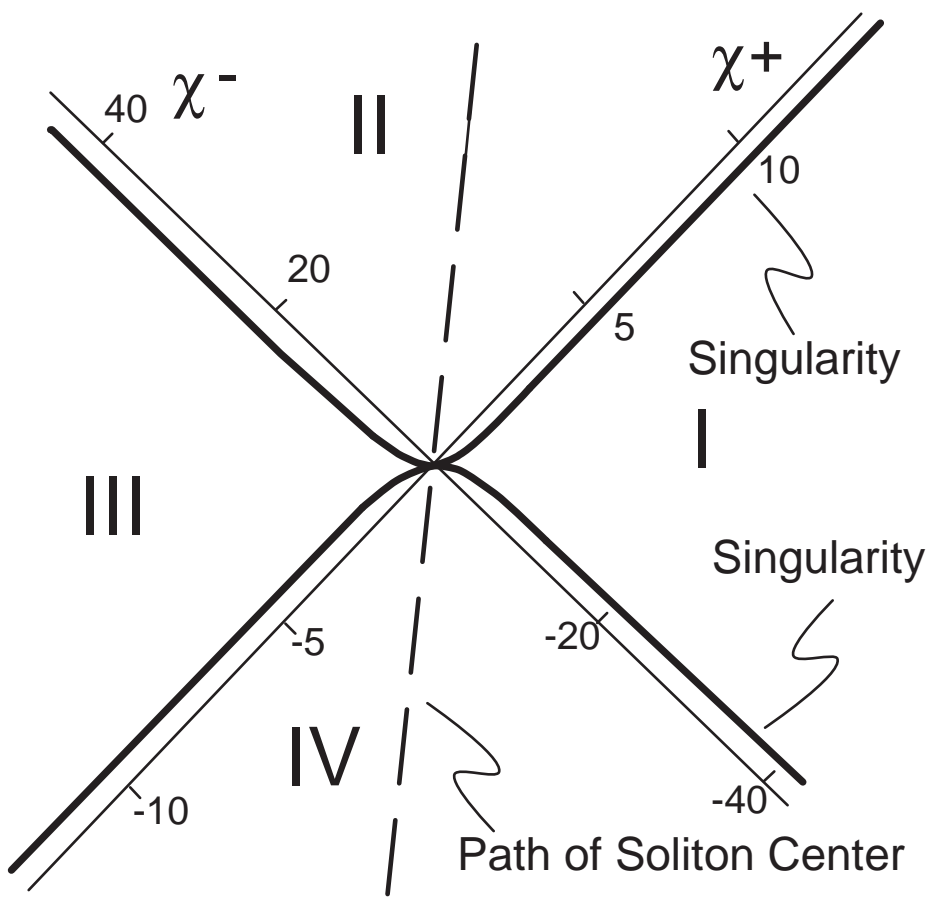
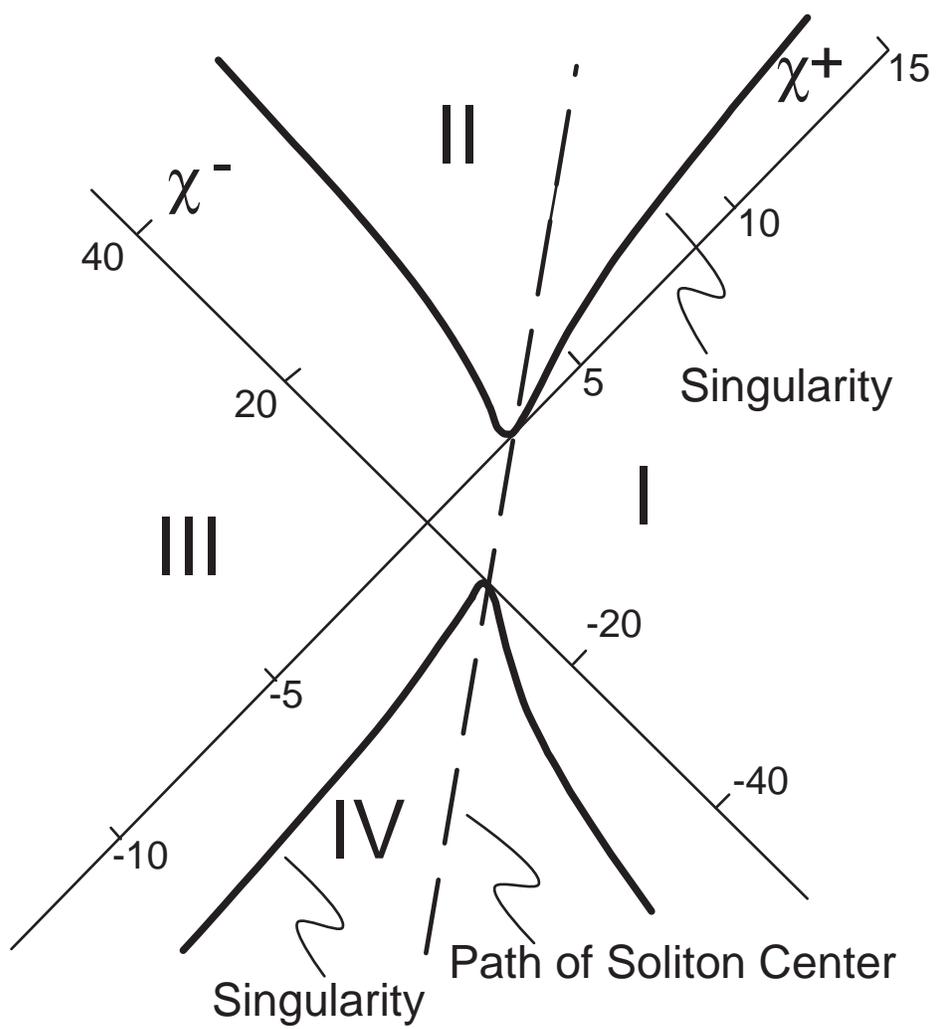
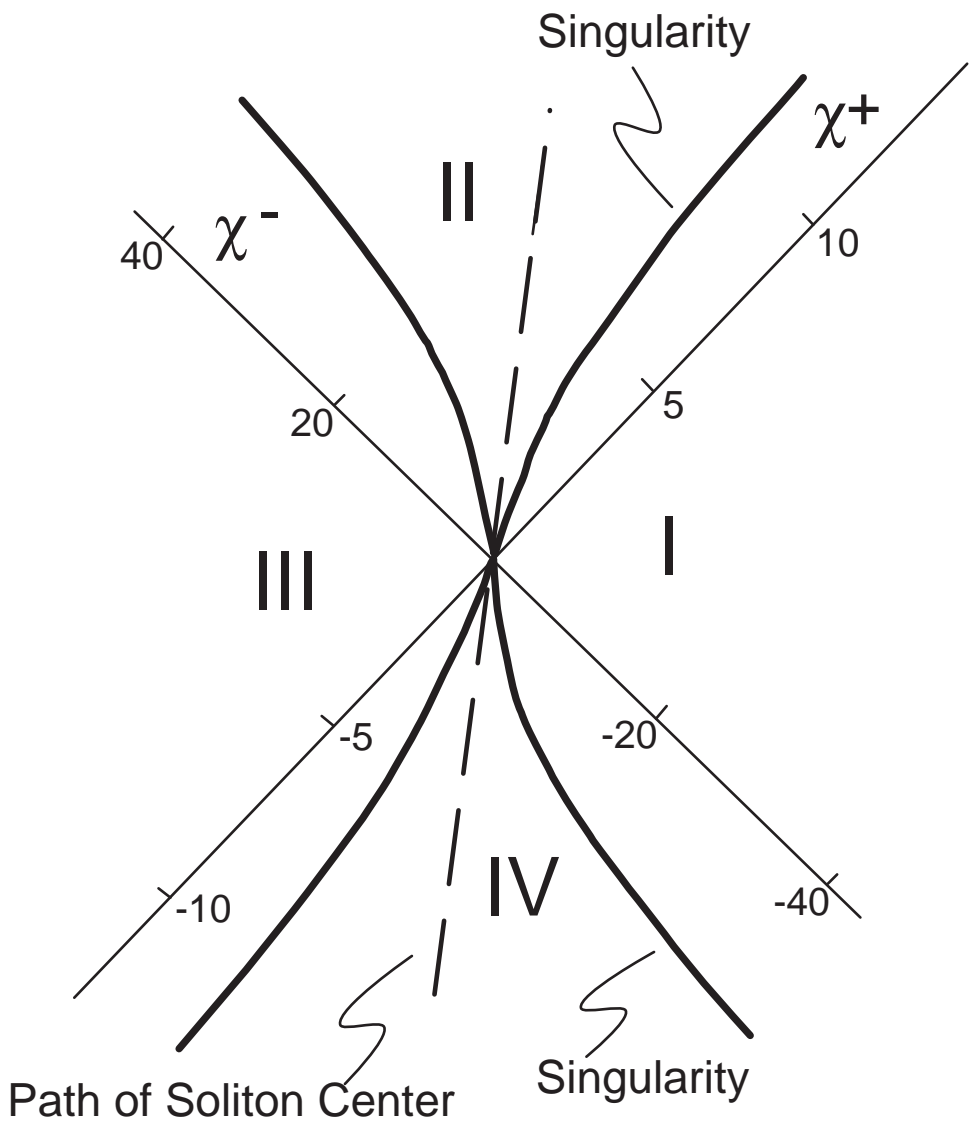
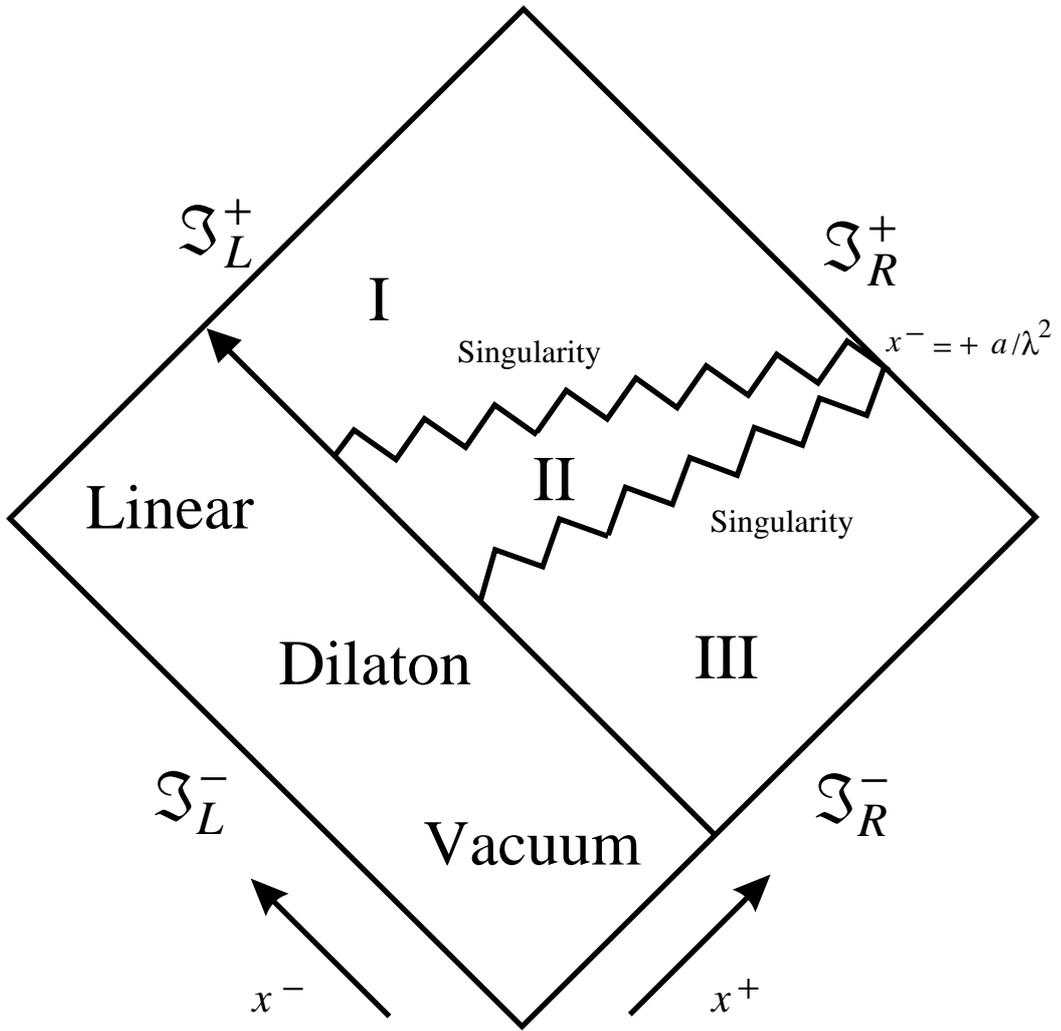
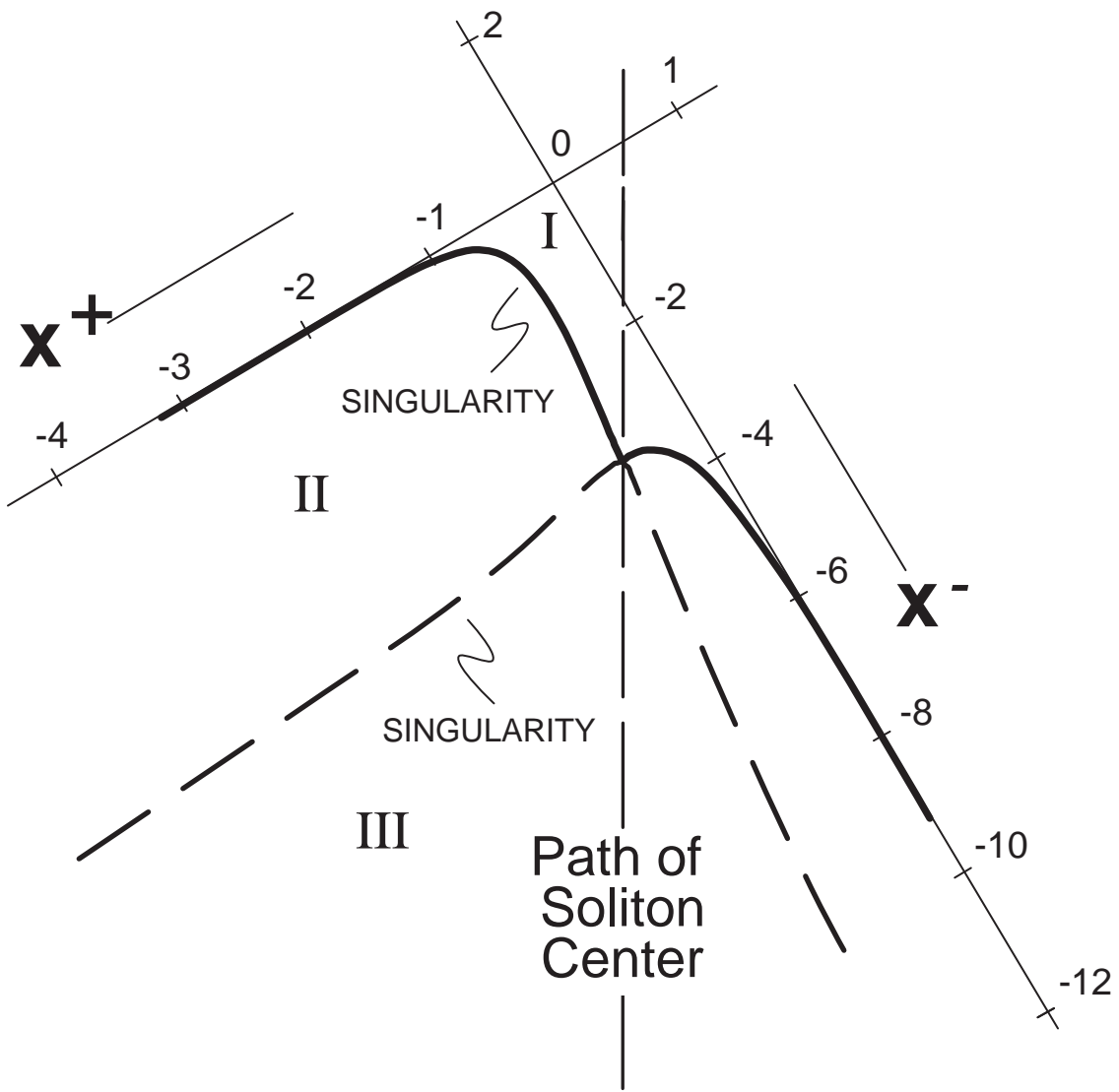
\end